\documentclass[twocolumn]{aastex63}

\usepackage{comment}
\usepackage[figure,figure*]{hypcap}
\usepackage{xcolor}

\usepackage{graphicx,natbib,placeins,amsmath}
\usepackage{multirow}

\newcommand{\teff} {\ensuremath{T_{\rm eff}}}

\newcommand{\Msun}{\ensuremath{M_{\odot}}}

\received{November 16, 2019}
\accepted{\today}
\submitjournal{ApJ}

\shorttitle{Orion sub-stellar binaries}
\shortauthors{Strampelli et al.}

\begin{document}

\title{HST survey of the Orion Nebula Cluster in the H$_2$O 1.4~$\mu$m absorption band:\\
III. The population of sub-stellar binary companions}

\correspondingauthor{Giovanni M. Strampelli}
\email{strampelligiovanni@jhu.edu}

\author[0000-0002-1652-420X]{Giovanni Maria Strampelli}
\affiliation{Johns Hopkins University, 3400 N. Charles Street, Baltimore, MD 21218, USA}
\affiliation{Space Telescope Science Institute, 3700 San Martin Dr, Baltimore, MD 21218, USA}
\affiliation{Department of Astrophysics, University of La Laguna, Av. Astrofísico Francisco Sánchez, 38200 San Cristóbal de La Laguna, Tenerife, Canary Islands, Spain}

\author[0000-0003-3184-0873]{Jonathan Aguilar}
\affiliation{Johns Hopkins University, 3400 N. Charles Street, Baltimore, MD 21218, USA}

\author{Laurent Pueyo}
\affiliation{Space Telescope Science Institute, 3700 San Martin Dr, Baltimore, MD 21218, USA}

\author{Antonio Aparicio}
\affiliation{Department of Astrophysics, University of La Laguna, Av. Astrofísico Francisco Sánchez, 38200 San Cristóbal de La Laguna, Tenerife, Canary Islands, Spain}
\affiliation{Instituto de Astrofísica de Canarias, C. Vía Láctea, 38200, San Cristóbal de La Laguna, Tenerife, Canary Islands, Spain}

\author[0000-0002-5581-2896]{Mario Gennaro}
\affiliation{Space Telescope Science Institute, 3700 San Martin Dr, Baltimore, MD 21218, USA}

\author{Leonardo Ubeda}
\affiliation{Space Telescope Science Institute, 3700 San Martin Dr, Baltimore, MD 21218, USA}

\author[0000-0002-9573-3199]{Massimo Robberto}
\affiliation{Johns Hopkins University
3400 N. Charles Street
Baltimore, MD 21218, USA}
\affiliation{Space Telescope Science Institute, 3700 San Martin Dr, Baltimore, MD 21218, USA}

\begin{abstract}
We present new results concerning the sub-stellar binary population in the Orion Nebula Cluster (ONC). Using the Karhunen-Lo\`{e}ve Image Projection (KLIP) algorithm, we have reprocessed images taken with the IR channel of the Wide Field Camera 3 mounted on the \textit{Hubble Space Telescope} to unveil faint close companions in the wings of the stellar PSFs. Starting with a sample of  1392 bona-fide not saturated cluster members, we detect 39 close-pairs cluster candidates with separation $0.16''-0.77''$. The primary masses span a range M$_p$ $\sim 0.015-1.27$  M$_{\sun}$ whereas for the companions we derive M$_c$ $\sim 0.004-0.54$  M$_{\sun}$. Of these 39 binary systems, 18 were already known while the remaining 21 are new detections. Correcting for completeness and combining our catalog with previously detected ONC binaries, we obtain an overall binary fraction of $11.5\% \pm 0.9\%$. Compared to other star forming regions, our multiplicity function is $\sim 2$ smaller than e.g. Taurus, while compared to the binaries in the field we obtain comparable values.
We analyze the mass function of the binaries, finding differences between the mass distribution of binaries and single stars and between primary and companion mass distributions. The mass ratio shows a bottom-heavy distribution with median value of $M_c/M_p \sim 0.25$. Overall our results suggest that ONC binaries may represent a template for the typical population of field binaries, supporting the hypothesis that the ONC may be regarded as a most typical star forming region in the Milky Way.
\end{abstract}
\keywords{binaries --- stars: pre-main sequence --- stars: low-mass --- open clusters and associations: individual (Orion Nebula Cluster)}

\section{Introduction}
Binary stars are coeval pairs of stars born in the same environment, with the same metallicity, but with different mass. Understanding their properties provide us with key information on stellar evolution, from the early phases of star formation to the most violent phenomenology that may characterize the final moments of their life. In what concern young systems,  knowing the effective temperature and absolute luminosity of a pair can constrain theoretical models developed to predict isochrones and evolutionary tracks on the HR diagrams during the Pre-Main Sequence phase \citep{Gennaro2012,Stassun2014}. Ignoring the presence of binaries, on the other hand, represents a nuisance that may affect the statistical analysis of the same HR diagrams \citep{Jerabkova2019}.

The distribution and frequency of binary systems with a substellar companions has been the object of several studies \citep[see e.g.][review and reference therein]{Duchene2013}. 
In principle, very low-mass companions \citep[down to the deuterium burning limit,][]{Spiegel2011} might form like stars through early fragmentation and gravitational collapse of a common pre-stellar core, or like planets in a circumstellar disk, reaching their observed wide orbits through migration or scattering. Characterizing the population of low-mass companion can thus shed light on the mechanism of star and planet formation at the lower and upper boundary, respectively, of their mass range.

Since substellar objects are unable to sustain hydrogen fusion in their cores and quickly fade away becoming undetectable, young stellar clusters in the solar vicinity are ideal for large statistical studies. 
Using direct imaging techniques, the main observational challenge is that objects potentially resolved may be hidden under the extended Point Spread Function (PSF) wings of the primary. No-detections only provide upper limits on the companion frequency within a wide range of mass and semi-major axis (SMA). To probe beyond these limits, image processing techniques that remove the PSF while preserving the flux of the companion have been developed.

The key element in performing PSF subtraction is having an accurate template for the PSF itself. In 1-to-1 PSF subtraction, also called Reference Differential Imaging, a single reference PSF is directly subtracted from the science image. For the two PSFs to match, reference and target images should be acquired maintaining the same instrument configuration, in the same part of the sky, and as close in time as possible. This helps reducing changes in the PSF due to variations resulting from e.g. the unstable thermal environment in a low-earth orbit environment, or instrument flexures and variable atmospheric conditions on the ground. In practice, if only one reference PSF is available, the results of the subtraction will always be subject to a variety of systematic and random differences between the reference and science images. To reduce the impact of using a particular realization of the reference PSF on the subtraction residuals, it is advantageous to combine multiple PSFs. A variety of observing strategies and algorithms have been developed in order to optimally combine multiple reference PSF images \citep[e.g.][]{Marois2014}. Eventually, in the case of a positive detection, finding a faint object in the immediate vicinity of a star does not provide conclusive evidence of a physical association. Complementary information, such as common proper or parallactic motion, is needed to disentangle real pairs from random alignments. Lacking multiple epoch data, the presence of photospheric features characteristics of young low mass objects may provide strong indication for real binary systems.

In this paper we presents the results of a search for substellar companions in the Orion Nebula Cluster (ONC) based on data obtained with the Hubble 
Space Telescope (HST). The ONC is ideal for this type of investigation: it is massive enough ($\sim 2000 M_{\sun}$) to provide us with a rich sample of targets 
and sufficiently nearby ($\simeq 400$ pc; \citealt{Kuhn2019}) that the the angular scale of a WFC3/IR pixel,  0.13'', corresponds to a physical separation of $\simeq 50\, \mathrm{AU}$, i.e. the distance of Pluto to the Sun at aphelion. 

Our strategy is based on reprocessing standard wide-field imaging data with advanced PSF subtraction techniques, namely the KLIP algorithm
\cite{Soummer2012}, fully exploiting the exquisite stability of the HST.
In particular, we have used a dataset consisting of images obtained with the IR channel of the Wide Field Camera 3 (\textit{HST}/WFC3) through a pair of filters tailored to measure the depth of the $1.4\,\mu m$ $\mathrm{H_{2}O}$ absorption feature: F139M (in band) and F130N (adjacent, line-free continuum). 
In the first paper of this series (Robberto et al., 2020, ApJ submitted, hereafter Paper~I with corresponding catalog of sources: Catalog~I) we have shown that the presence of the water absorption feature in the atmosphere of low luminosity sources can be used to separate the substellar cluster population of the Orion Nebula Cluster (ONC) from background stars and galaxies. 
The flux decrease in the F139M filter relative to the nearby F130N continuum produces a negative (blue) m$_{130}$-m$_{139}$ color index 
highly sensitive to the effective temperature down to $\teff\simeq 2800$~K ($\sim 0.06$~M$_{\Sun}$); below this value the absorption feature remains strong but with a weaker dependence on the effective temperature, reaching m$_{130}$-m$_{139}$ $\simeq -0.5$ at temperature $\teff\simeq 2200$~K  ($\sim 0.01$~M$_\Sun$).

The possibility of discriminating low-mass objects from the population of reddened field stars has allowed Gennaro \& Robberto (2020, ApJ submitted, hereafter Paper~II) to investigate the shape of the initial mass function of ``field'' cluster members down to planetary masses.  
Catalog~I, however, only reaches  separations as small as $0.8 ''$ (320~AU), inside of which the search for binary candidates is hampered by PSF blending. By applying the KLIP algorithm and advanced statistical analysis to discard false positive detections,
we are able to provide a new, comprehensive picture of binarity in the ONC from $70$ to $310$ AU

In Section \ref{section:dataset}, we summarize the main characteristics of the dataset.
In Section \ref{section:method} we present our methodology whereas
in Section \ref{section:results} we present the result of our search.
We discuss the main properties of our sample in Section \ref{section:discussion}, while in Section \ref{section:conclusion} we summarize our findings.

\begin{figure}[t!]
\begin{center}
\includegraphics[width=0.49\textwidth]{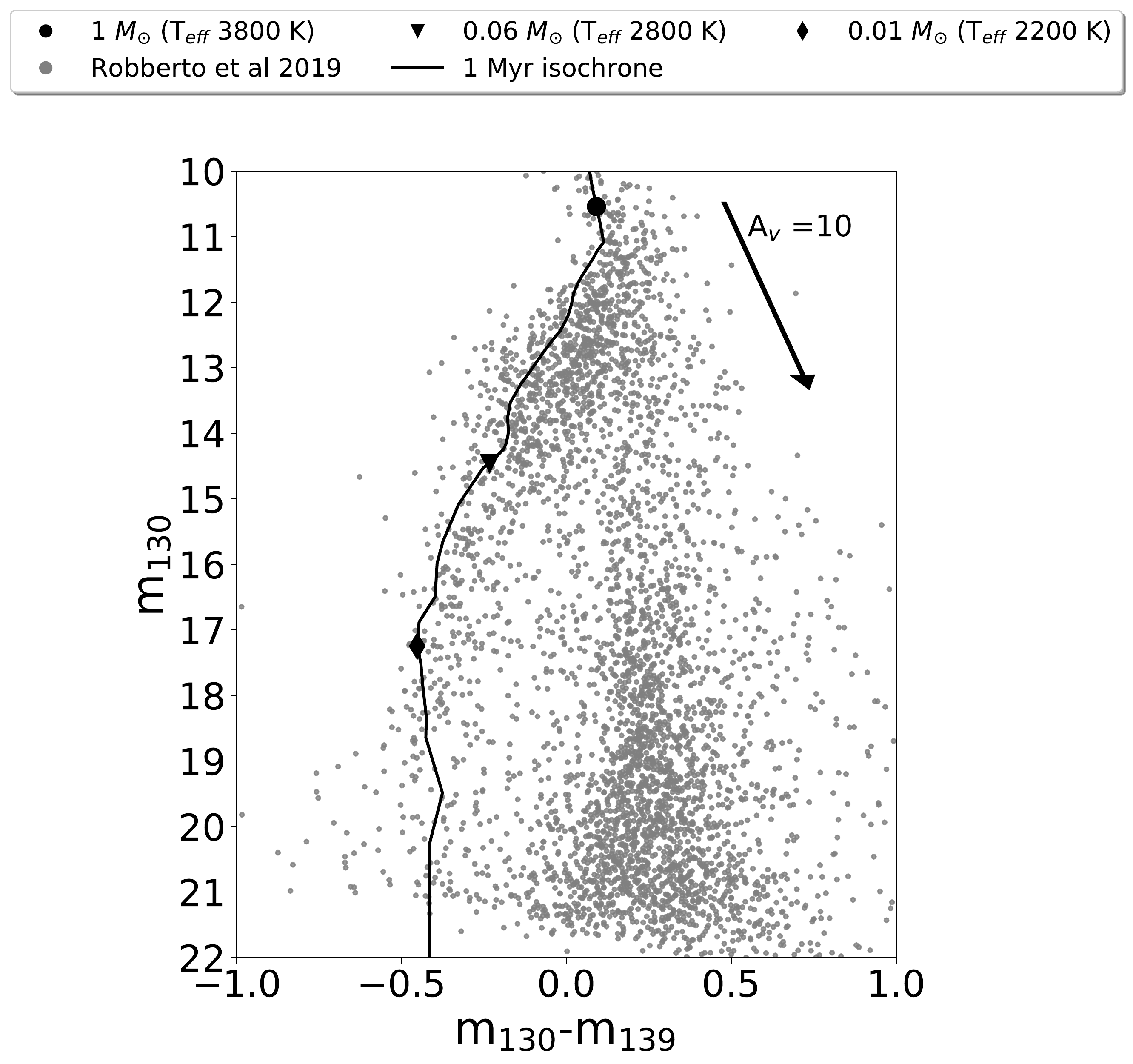}
\caption{Color magnitude diagram in filters F130N, F139M for all sources detected in the ONC field. The black line is a 1 Myr isochrone with three characteristic masses, from bottom to top $M_{\sun} = 0.02,0.08$ and $1$. \label{Fig:plain_cmd}}
\end{center}
\end{figure}

\section{Dataset}\label{section:dataset}
The Cycle 22 \textit{HST} Treasury Program "The Orion Nebula Cluster as a Paradigm of Star Formation" (GO-13826,  P.I.  M.  Robberto) aims at reconstructing the low-mass IMF down to $\sim 5 M_{Jup}$ in the ONC. Paper~I presents the survey strategy, sensitivity limits and completeness analysis, leading to a census of the stellar and substellar population in the ONC down to few Jupiter masses in the F130N and F139M filters.  The 208 images taken in each filter produce wide field mosaics covering an area of $\sim 1/6$ of a square degree. The number of \emph{unique}  sources, either ONC members or background stars and galaxies,  is 4504 but in this paper we reprocess the full dataset of more than $\sim 8700$ source detections, as the mosaicing strategy allowed detecting the same sources during multiple visits. 
Figure \ref{Fig:plain_cmd} shows the color-magnitude diagram for all 4504 sources, with the clear separation between the cluster population at the top and left side of the diagram, and the background sources at bottom right, with positive $m_{130}-m_{139}$ color. A 1~Myr isochrone, adapted from the BT-Settl model to correct for the discrepancy between the model and the data, is overplotted  in red color up to a mass  $M\lesssim 0.75 M_{\sun}$ (see Paper~I for a description of the models and of their semi-empirical calibration). 
For masses $\geq 0.75 M_{\sun}$ we departed from the BT-Settl model, adopting instead the MESA isochrones and Stellar Tracks (MIST) for the WFC3 IR channel in our F130N and F139M filters \citep{Dotter2016,Choi2016}. 



\section{Data Analysis}
\label{section:method}
\subsection{Catalogs of reference and target stars}
As reported in Paper~I, saturation in the F130N filter starts at $m\simeq 10.9$ while the noise floor is at $m\simeq22$, setting the magnitude limits of the primaries and companions we are able to analyze. 

Our input catalog of targets contained 8210 individual detections (of which 4220 unique) with m$_{130}$ magnitudes in the range 10.9 to 22, about 50\% of them corresponding to repeated detections of the same sources.

Our PSF subtraction technique requires a reference catalog of sources uncontaminated by astrophysical or instrumental noise. We create it from our sample, perform several clean-up steps:
\begin{enumerate}
    \item \emph{Visual binaries removal}: We remove from our catalog 157 unique pairs, for a total of 623 total entries with a neighbor closer than 1.5'' projected distance according to the Catalog~I. In this way we avoid contamination from nearby neighbours whose PSFs wings may affect the region searched for low-mass companions.
    \item \emph{Bad pixel removal}: The dataset of full-frame WFC3 images is cleaned from cosmic-rays event in the early stages of standard data processing thanks to the non-destructive sampling of the accumulating signal. Static bad pixels are also flaggeed by the pipeline. However, we perfom an independent check by stacking the images and applying a 10$\sigma$ threshold to the distribution of median pixel values. We didn't find any detection with a flagged pixel closer than $\sim 0.8''$ in any visit. 
    \item \emph{ACS catalog matching}: \textit{HST}/ACS survey of  \cite{Robberto2013} provides a high-resolution morphological classification of the sources in the ONC. By cross-matching the ACS catalog  with our list of WFC3 detections, we discard all objects flagged as non-stellar, i.e. silhouette disk, proplyds, sources with evidence of jets/photoionization, Herbig-Haro objects, resolved galaxies. We discarded a total of 222 unique objects for a total of 458 entries from the catalog
\end{enumerate}

Applying these selection criteria we end up with with a catalog of 7129 individual sources, counting multiple observations of the same object separately.

The next step is to create ``postage stamps'' centered on each source and perform the PSF subtraction inside this area. In setting our $11\times11$ ($1.5''\times1.5''$) pixel stamp size we
consider the following factors:
\begin{itemize}
    \item the area  must be large enough to contain the bright wings of the PSF, for sources matching our assumed range of magnitudes;
    \item the area must have enough pixels to provide a meaningful noise calculation. Detections of close companions are affected by small number statistics and a correction to the  estimated contrast and SNR has to be applied \citep{Mawet2014}. The following argument shows that the correction is very small for an 11$\times$11 stamp. The number of $\lambda/D$ resolution elements per pixel for WFC3 in the F139M filter is close to 1, i.e. WFC3-IR is significantly undersampled. Therefore, a $11\time11$ pixel stamp contains about the same number of resolution elements. The correction factor to the SNR is given by $\left(\sqrt{1+1/n}\right)^{-1}$, which for $n=121$ is 0.996. 
    Therefore, the sample size does not represent a significant source of uncertainty vs. other noise sources, e.g. photon or read noise.
    \item the area must be small enough so that tiles do not overlap; having rejected from our catalog objects with a nearest companion closer than 1.5", this results in a tile half-size of 0.7". With a WFC3 pixel scale of $0.13''\,\mathrm{pixel}^{-1}$, the tile half-size translates to a radius of approximately 280~AU distance from a point source in the ONC.
\end{itemize}


\subsection{PSF subtraction}\label{subsection:PSF subtraction}
Accurate PSF subtraction depends strongly on the quality of the reference PSF, a task greatly simplified by the stability of the Hubble Space Telescope which has enabled the compilation of libraries of PSF models for reference differential imaging \citep[e.g.][]{Choquet2014}. 
Still, for the most accurate PSF subtraction one has to deal with the field distortion of WFC3 and the small but not negligible time-dependence of the HST focus. These effects make the PSF both spatially and time dependent. Our strategy is especially well-suited for handling both effects.

It consists in dividing the field of view into 100 equal cells, each cell small enough to neglect local PSF distortion but large enough to build a local PSF library containing enough stars to build an accurate model. 

For each cell, PSF subtraction is then performed as follows: 
\begin{itemize}
    \item the postage stamp for all stars in the cell are stacked together into a single data cube;
    \item iterating through the data cube, each stamp is assumed as the science image;
    \item a reference model of the PSF for subtraction is  constructed selecting from the remaining postage stamps those with a photometric error $ \sigma_{F130N} \leq 0.01$;
    \item the PSF of the target star was in then removed using the Karhunen-Lo\`{e}ve Image Projection (KLIP) algorithm \citep{Soummer2012}. 
\end{itemize}

For each target, we chose the number of modes which simultaneously minimize the standard deviation of the residual image while maximizing the counts of the brightest residual pixel. 

To build a preliminary catalog of candidate binaries, we analyze the position of the brightest pixel of the residual images of each target. To be labeled as a candidate detection, at this early stage, we require that: 
\begin{itemize}
    \item the pixels with the highest flux in each residual must be within one pixel in both filters and in all available visits when the source is observed with different telescope orientations;
    \item to candidate must be detected in at least two different KLIP modes.
\end{itemize}
The one pixel distance (rather than zero) is needed to take into account possible misalignments of the center of the stars in the reference library, due to the undersampled PSF and lack of dithering in the survey. This reflects in an accuracy of our separation estimates of about 1/2 pixel, i.e. 0.07" or 28~AU at the distance of the ONC.


\subsection{Cluster and Background candidates}
\label{subsection:CB candidates}
The inspection of the residuals immediately after PSF subtraction reveals a large number of candidate companions, but further down-selection has to be applied to reject sources that presumably do not belong to the ONC. 
To separate cluster stars from background sources we use the position of the stars on the CMD. 
As shown in Paper I, the pair of filters chosen for this survey is sensitive to the depth of the $1.4\,\mu m$ $\mathrm{H_{2}O}$ absorption band. This temperature-sensitive feature is prominent in the atmosphere of M-type stars and brown dwarfs, down to planetary-mass objects and can be than used to separate the substellar cluster population of the ONC from background stars and galaxies.
Following Paper~I we consider a source to be a ONC member if it lies in the area delimited by the 1 Myr isochrone introduced in Section \ref{section:dataset}, reddened by $A_V = 10$ mag. 
Any companion candidates bluer (redder) than this isochrone is labeled as cluster (background). In Paper~I we found good agreement between this simple approach and a more rigorous Bayesian statistical treatment. At the end of this process we obtained 2797 multiple visits cluster sources, with 1392 \emph{unique} targets for our KLIP PSF subtraction algorithm.


\subsection{Companion Photometry}
\label{subsection:Photometry and Mass estimation}
Since the WFC3/IR PSF is highy undersampled,
after PSF subtraction we expect most of the flux from a faint candidate companion to be contained within a few pixels. Thus, to derive the total flux one has to apply a large and rather uncertain aperture correction. To evaluate it, we analyze a sample of isolated bright stars in our catalog comparing their flux around the brightest pixels with their total flux.
This analysis shows that about 1/3  of the flux is contained within the brightest pixel and  $\sim 60\%$ within the four adjacent brightest pixels. The distribution of relative fluxes for the four brightest pixels is narrower than the distribution for the single pixel. Therefore, we perform our photometry of the companions using a 4-pixel aperture,
deriving the aperture correction to the total flux through comparison with the Catalog~I PSF photometry. Specifically, 
for each isolated source in Catalog~I we built a square 2x2 pixel mask placed so that one pixel always coincides with the brightest pixel of the original image. After probing the 4 possible mask positions, we record the maximum value of the total counts as c$_{4p}$. The magnitude for each primary is then calculated as:

\begin{equation}
\label{eq:m4p}
m_{4p} = -2.5 log_{10}(c_{4p}) + C
\end{equation}
where $C$ is a normalization factor between the 4-pixel photometry and PSF photometry ($C_{m_{130}} =21.35\pm0.049$, $C_{m_{130}-m_{139}}=-0.002\pm0.031$).
We then determine $C$ as the mean of the difference between the PSF photometry and the 4-pixel photometry of each primary:
\begin{equation}
\label{eq:C}
<m_{PSF}-m_{4p}> = C
\end{equation}
Measuring c$_{4p}$ for each detected companion and using equations \ref{eq:m4p} and the value of $C$ from equation \ref{eq:C}, we determine the magnitudes of our candidate companions. Our estimate of the total uncertainty takes into account the uncertainty on the counts of the candidate, on the background counts in the 4-pixel aperture, and on the estimated conversion factor between the PSF and the 4-pixel system (the standard deviation of the sample we used to evaluate the conversion factor).

Having determined the photometry for each candidate, a new selection is applied keeping all the cluster pairs with companion magnitude in the range $10.9 \geq$ mag$_{130}$ $\geq 22 $ 
(following a similar approach as the one mentioned in Section \ref{subsection:CB candidates})
and with absolute value of the m$_130$-m$_139$ color $\leq 1$ to reject noisy outliers. This results in a preliminary selection of 145 cluster candidate binaries. 



\subsection{Real vs. false positive detections}\label{subsection:Candidate selection}
To assess our ability to separate plausible candidate from instrument induced false positive detections, we perform an extensive set of simulations to determine the Receiving Operating Characteristic (ROC) curves (see Appendix \ref{app_section:ROC curves} for an explanation of ROC curve construction) for each binary configuration in our preliminary catalog. A configuration is specified by three parameters a) brightness of the primary, b) contrast between primary and companion, anc c) separation and KLIP mode used during the PSF subtraction phase. We use the ROC curves to derive three other quantities we can use to make the following selections on our candidates:
\begin{itemize}
    \item the Area Under the Curve (AUC) of the ROC: the AUC provides us with a good indication of how well the distribution of the true positive rate (TPR, i.e. detection of companions injected in our simulations) is separated from the distribution of the false positive rate (FPR, i.e. detection of noise peaks that may have been erroneously determined to be companions). A AUC curve of 0.5 indicates that there is no possibility of separating the two distributions, whereas an AUC=1 represents perfect separation. 
    An analysis of the results provided by the simulations led us to select a candidates only when the corresponding configuration provides an AUC $\geq 0.7$. 
    
    \item false positive probability and SNR threshold: as explained in Appendix \ref{app_section:ROC curves},  for each given configuration, the ROC curve is built sliding a SNR threshold across the TPR and FPR distributions. We can therefore invert this process: given the ROC curve for the certain configuration and having determined a limit to the probability for a detection to be a false positive, we find the corresponding SNR that we can use as a threshold for the detection. 
    Because each candidate is found using multiple independent detections (different filters and possibly different locations on the detector for each visit), we multiply the false positive probabilities of each detection ($FP'$) to obtain an overall false positive probability for the whole candidate ($FP$). In particular, if we assume $FP'$ to the same for each detection, it is:
    \begin{equation}
        FP=FP'^{(N_{f} \times N_{v})}
    \end{equation}
    where $N_{f}$ is the number of filters and $N_{v}$ is the number of visit for the candidate. 
    Inverting this relation we find $FP'$ as a function of $FP$. 
    Having set $FP'$, we can find the corresponding SNR threshold from the ROC. With 1392 primaries to be searched, assuming an overall  false positive probability $FP = 0.2\%$ for each candidate, we expect $\sim 3$ false positive detection in our final catalog of binaries. We have verified that this probability value represents an optimal trade-off. A further reduction, i.e. a more aggressive reduction of false positives, would imply higher detection thresholds that would lead to reject strong previously known true detections. Viceversa, relaxing the threshold would cause a large increase in the number of false positives beyond the acceptable rate of 50-100 smaller than the expected detection signal (as a point of reference the expected binary fraction is ~10-20 percent as per Kraus and Duchene)
    
    \item ratio of true positives over false positives (R): for each candidate detection we binned the TPR and FPR distributions in bins of 0.5 SNR and we evaluate the ratio of true positive over false positive in the same bin corresponding to the candidate SNR detection. This parameter give us an indication about how common the candidate SNR is in the distribution of false positive and true positive. Because each candidate results from multiple detections, we keep only candidates with with an $R_{median} \geq 3$. 

\end{itemize}

\begin{figure}[!t]
\begin{center}
\includegraphics[width=0.49\textwidth]{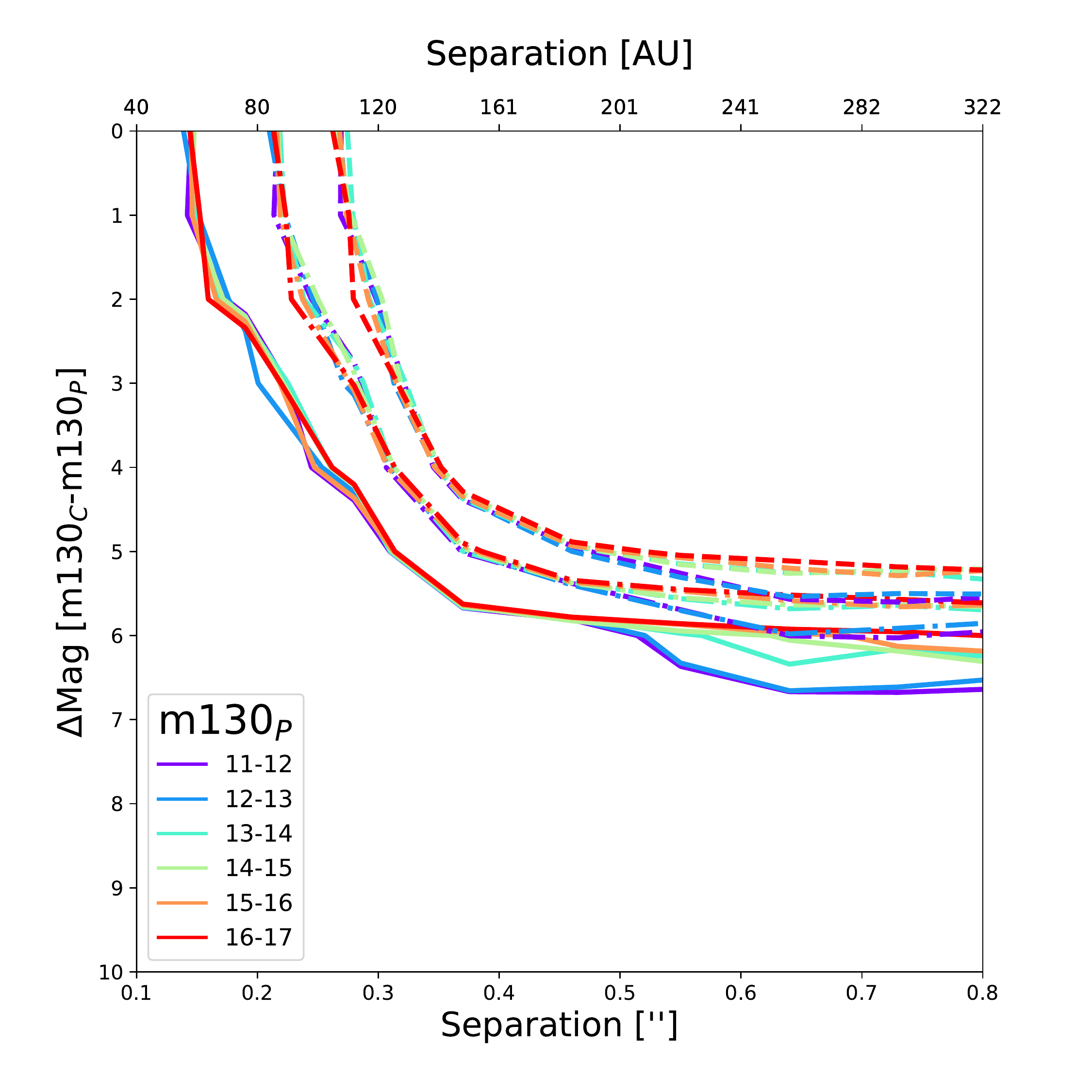}
\caption{Averaged contrast curve over different visits and KLIP modes for each magnitude bin of the primary star and delta magnitude between companion and primary and projected separation. the three families of curves correspond to \emph{completeness} $\mathcal{C}=0.1, 0.3, 0.5$ \label{Fig:Contrast_curves}}
\end{center}
\end{figure}

As a by-product of our simulations, we also obtain the amount of flux lost due to over-subtraction (see \citealt{Pueyo2016} and reference therein), deriving the correction to apply to the photometry
of our candidate companions, with the relative errors. 
Moreover, from the distributions of TPR and FPR we can also evaluate the contrast curves as a function of the magnitude of the primary, contrast and separation. Averaging all data we obtain the contrast curves shown in Fig. \ref{Fig:Contrast_curves}.

The preceding analysis is not designed to distinguish between true companions and other astrophysical sources of false positives. These include residual contamination from nearby stars and light emitted by circumstellar material. Detector persistence may cause "ghosts" of very bright stars into the subsequent exposures, but they also appear as extended structures that can be easily identified and generally decay within one visit (see Paper~I).
This is why to conclude this candidate selection we visually inspect all our selected candidates looking for extended residuals.

\subsection{Companion Mass Determination}
\begin{figure}[t!]
\begin{center}
\includegraphics[width=0.49\textwidth]{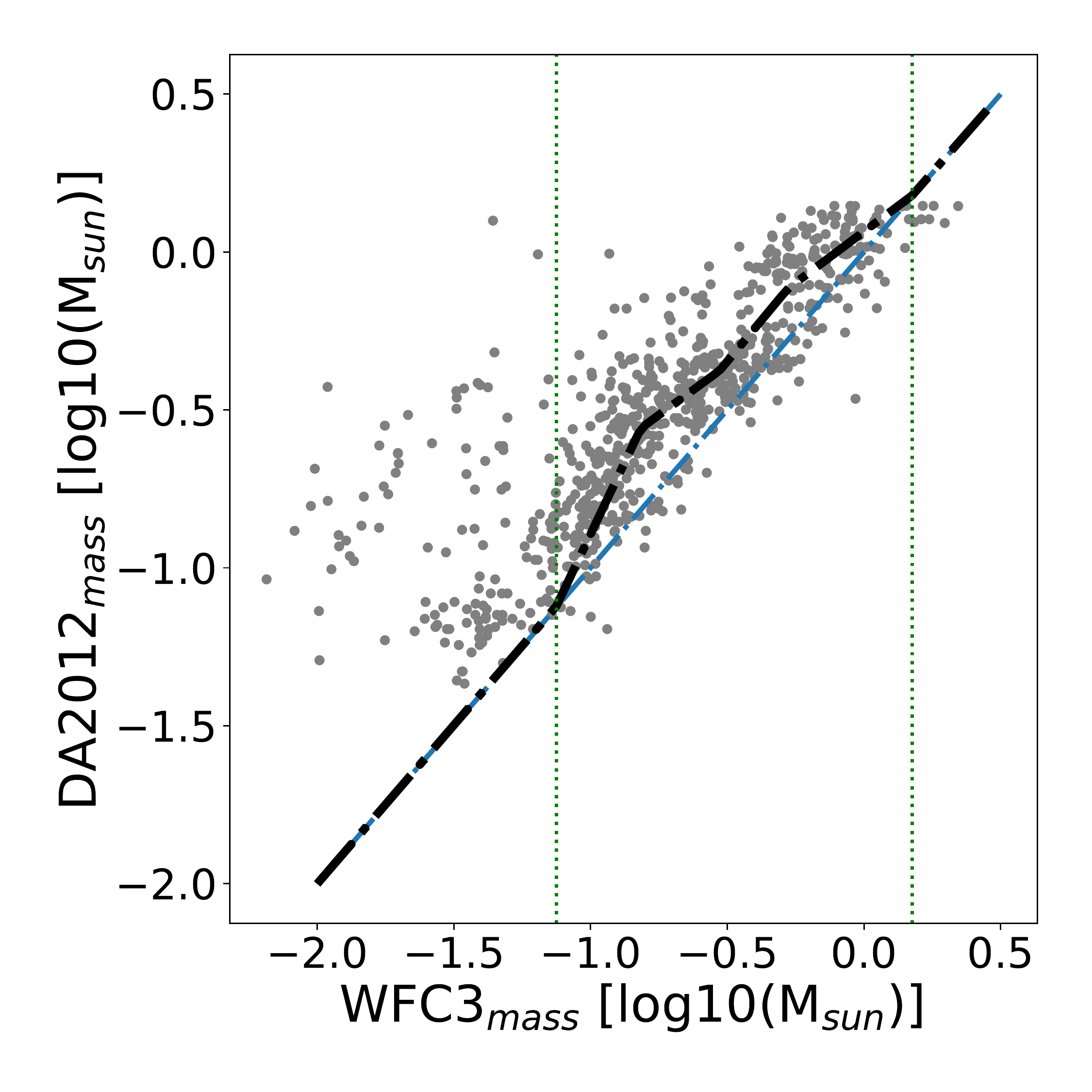}
\caption{\citeauthor{DaRio2012} DR2012 vs. WFC3 masses (grey points). The two green dotted lines mark the value for WFC3$_{mass}=0.075\:M_{\Sun}$ and $1.5\:M_{\Sun}$. The blue dotted line shows the locus of point where DR2012$_{mass}$=WFC3$_{mass}$ \label{Fig:DaRioF130mass}, while the black dotted line shows the final spline fit of the data.}
\end{center}
\end{figure} 

To estimate the mass of our substellar companions we start with an analysis of the primaries and isolated ONC stars. 
Figure \ref{Fig:DaRioF130mass} shows the comparison between the masses estimated by \citet{DaRio2012} using the \cite{Baraffe1998} evolutionary models (DR2012$_{mass}$)  and the masses obtained from our de-reddened WFC3 photometry and the 1 Myr isochrone (WFC3$_{mass}$, grey points). We use the value of A$_{V}$ determined by \citeauthor{DaRio2012} when available, otherwise we use the A$_{V}$ estimate from Paper~I , with negative A$_V$ values are set to A$_V=0$.
In the range of WFC3$_{mass}$ between $0.075 - 1.5 M_{\Sun}$ (vertical lines in the plot) we observe good correlation with some systematic difference between the two mass estimates.
Below this range, the scatter increases, an indication of the difficulty of DR2012 optical survey in dealing with the reddest and faintest sources of their sample.
To reconcile the two datasets, 
we use an empirical isochrone, fitting the relation between the DR2012$_{mass}$ and the WFC3$_{mass}$ in the $0.075 - 1.5 \:  M_{\Sun}$ mass range with a spline function as follow:
\begin{itemize}
    \item we bin the distribution of F130N$_{mass}$ between $0.075 - 1.5 M_{\Sun}$. To have bins  perpendicular to the DR2012$_{mass}$=WFC3$_{mass}$ relation (blue line in Figure~\ref{Fig:DaRioF130mass}) we apply a rotation matrix to the data by an angle of 45 degree; 
    \item we apply a 3-sigma cut to the distribution of each bin to exclude outliers;
    \item we rotate back the data and we fit a spline matching the 1 Myr isochone outside the $0.075 - 1.5 M_{\Sun}$ WFC3$_{mass}$ range and the median green point otherwise (black dotted line in Figure~\ref{Fig:DaRioF130mass}).
\end{itemize}
In the substellar regime, instead, we only use our WFC3 data relying on the strong correlation between mass and stellar flux ($m_{130}$), as evidenced by the color-magnitude diagram (Figure~\ref{Fig:plain_cmd}).

Finally, to evaluate the mass of our candidate binaries, we assign the same $A_V$ values to both components and then evaluate the mass of the companion using the 
spline curve.

\subsection{Completeness limit}\label{subsection:Mass completeness curves}
\begin{figure}[!t]
\begin{center}
\includegraphics[width=0.49\textwidth]{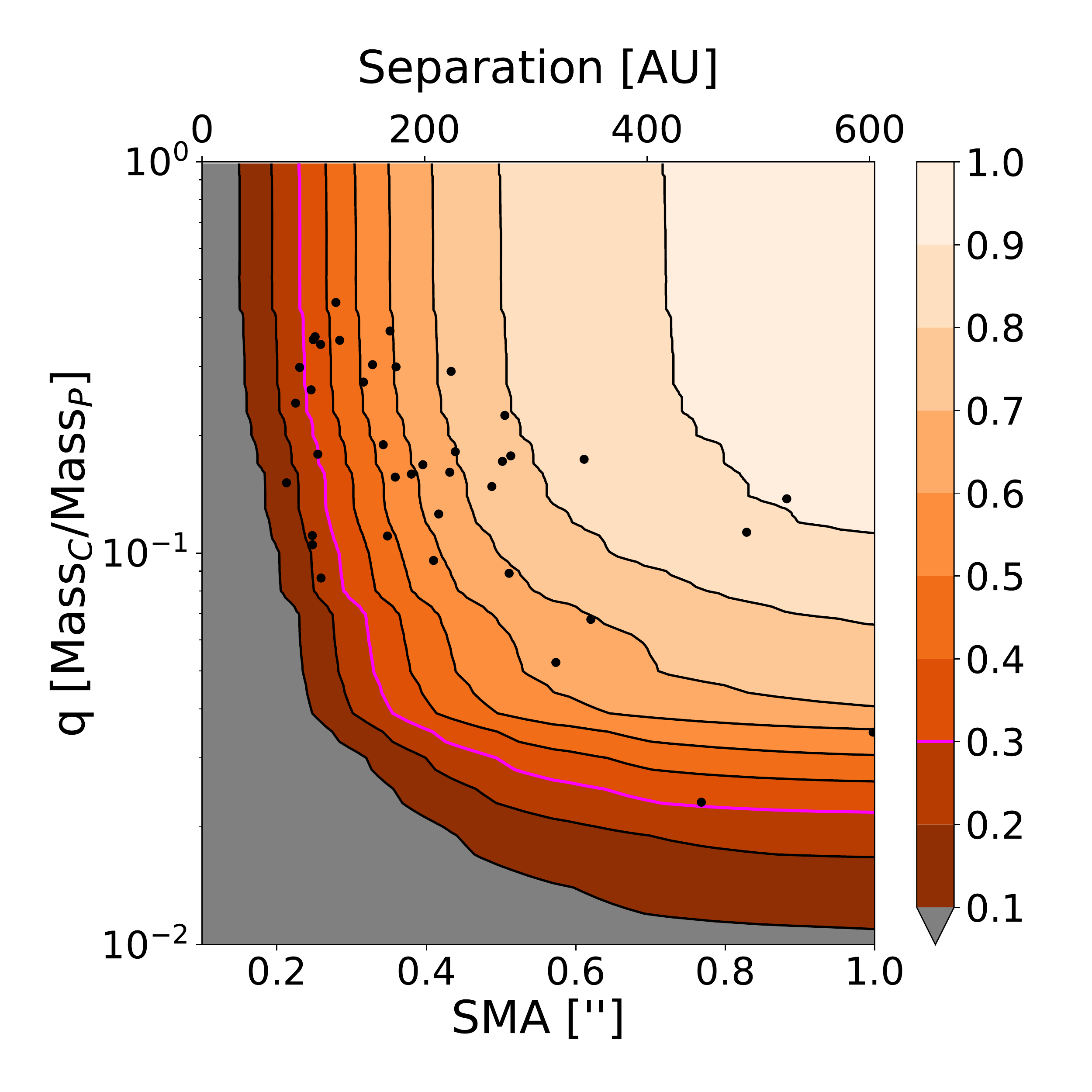}
\caption{Mass ratio completeness curves estimated as a function SMA with the color coding the value of the completeness. The black dots marks position of each detection on the plot. The magenta line marks the 30\% limit below which we only choose candidate with two or more visits. \label{Fig:CoC}}
\end{center}
\end{figure}

The completeness of our survey depends on the mass of the primary, the mass ratio of potential candidate and their separation, i.e. the projected SMA. This function, marginalized over the mass of the primaries, can be represented by a set of completeness curves for the mass ratio of the candidate and separation. 
Completeness as a function of the magnitude of the primary, companion, and visual separations can be obtained by direct inspection of the family of ROC curves discussed in Section~\ref{subsection:Candidate selection}. It can then be converted in a completeness as a function  of primary mass, mass ratio and deprojected orbital SMA. This last step is carried out using the following procedure:
\begin{itemize}
    \item we interpolate over a finer grid both in mass ratio and separation. 
    \item following \cite{Brandt2014}, we integrate over all the possible semi-major axes (s) between 0 and 1.8 using a piecewise function $p(s)$:
    \begin{equation}
      p(s) \simeq 
        \begin{cases}
            1.3s \quad\quad\quad\quad\quad 0 \leq \: s \: \leq \: 1\\
            -\frac{35}{32}(s-\frac{9}{5}) \quad\quad 1 < \: s \: < \: 1.8
        \end{cases}       
    \end{equation}
\end{itemize}
We then use this completeness map to apply a final selection to our catalog of candidates to reject any detection with completeness smaller than 10\% or between 10\%-30\% and with only one visit (i.e. the most likely to be one of the few false positive we expect, since our FP analysis was carried out using single visits). At the end of this selection process, we obtaining a final catalog of 39 reliable cluster candidates binaries out of 1392 original cluster targets. 

Figure \ref{Fig:CoC} shows the final completeness curves as a function of separation in SMA, the black dots mark the position of our detections on the completeness map. The magenta line show the 30\% completeness cut we apply to our single visit detection, while the gay area shows the space of parameters in the plot where we always reject candidates because completeness is smaller than 10\%.


\section{Results}
\label{section:results}
\subsection{Catalog of KLIP-detected candidate cluster binaries }
The analysis described in Section \ref{section:method} provides us with a total of 39 candidate cluster binaries with separation in the range $1.26-5.9$ pixels ($0.16''-0.77''$), corresponding to about $66-309$ AU projected distance from the primary assuming a distance of $403$ pc \citep{Kuhn2019}. The primary masses range between  $0.015$  M$_{\sun}$ - $1.27$ M$_{\sun}$ while the companions are in the range $0.004$  M$_{\sun}$ - $0.54$ M$_{\sun}$.

Table~\ref{Tab:CBC_klip} shows the physical and photometric properties of the 39 candidates.
Column (1) shows the entry number in the catalog; 
columns (2) and (3) show the Right Ascension and Declination for Equinox J2000.0; columns (4) to (11) list the $m_{130}$ magnitude and the $m_{130}-m_{139}$ color with their relative uncertainties for both primary (P) and companion (C); columns (12) to (15) show the estimated mass from F130N photometry, with its  uncertainty, for both primary and companion in units of Solar mass. The last three columns 
list the position angle, the separation between primary and companion, and the distance of the system from the core of the cluster (identified by the position of $\theta^1$Ori-C).

In  Appendix \ref{app_section:Gallery of binaries} we present a gallery of postage stamps (Figure \ref{Fig:KLIP_gallery0}-\ref{Fig:KLIP_gallery1}) showing the  co-added images before and after KLIP subtraction for each candidate, the companions generally appearing as bright single pixels in each residual image due to the WFC3/IR sub-sampling. Each postage stamp has dimensions $2''\times2''$ and is rotated so that north is up and east to the left. 

\begin{longrotatetable}
\movetabledown=15mm
\begin{deluxetable*}{chhhccccccccccccccccc}
\tabletypesize{\footnotesize}
\setlength{\tabcolsep}{2.pt}
\tablewidth{0pt}
\tablecaption{\textit{Candidate Binaries catalog (KLIP)}
\label{Tab:CBC_klip}}
\tablehead{\colhead{ID} & \nocolhead{WFC3ID$_p$} & \nocolhead{WFC3ID$_c$} & \nocolhead{SimbadName} & \colhead{Ra$_p$} & \colhead{Dec$_p$} & \colhead{mag130$_p$} & \colhead{color$_p$} & \colhead{dmag130$_p$} & \colhead{dcolor$_p$} & \colhead{mag130$_c$} & \colhead{color$_c$} & \colhead{dmag130$_c$} & \colhead{dcolor$_c$} & \colhead{mass$_p$} & \colhead{mass$_c$} & \colhead{emass$_p$} & \colhead{emass$_c$} & \colhead{PA} & \colhead{Sep} & \colhead{SepOriC} \\ 
\colhead{(--)} & \nocolhead{(--)} & \nocolhead{(--)} & \nocolhead{(--)} & \colhead{(deg)} & \colhead{(deg)} & \colhead{(mag)} & \colhead{(mag)} & \colhead{(mag)} & \colhead{(mag)} & \colhead{(mag)} & \colhead{(mag)} & \colhead{(mag)} & \colhead{(mag)} & \colhead{(solMass)} & \colhead{(solMass)} & \colhead{(solMass)} & \colhead{(solMass)} & \colhead{(deg)} & \colhead{(arcsec)} & \colhead{(arcsec)} }
\startdata
0 & 112.0 & -1.0 & JW  39 & 83.65873105 & -5.461256454 & 12.7852 & 0.07241 & 0.06655 & 0.00258 & 14.79325 & -0.1649 & 0.06521 & 0.10838 & 0.265761 & 0.045624 & 0.007075 & 0.001164 & 11.04 & 0.39 & 630.53 \\
1 & 277.0 & -1.0 &  & 83.75715053 & -5.452742454 & 13.6095 & -0.18075 & 0.01708 & 0.0045 & 15.57923 & -0.38044 & 0.03513 & 0.06504 & 0.119621 & 0.031277 & 0.000737 & 0.000623 & 99.61 & 0.19 & 316.95 \\
2 & 326.0 & -1.0 & JW 290 & 83.77675985 & -5.451311859 & 12.7123 & -0.27704 & 0.01868 & 0.00997 & 13.52197 & -0.55827 & 0.05769 & 0.10461 & 0.280446 & 0.100214 & 0.002345 & 0.002273 & 280.28 & 0.19 & 268.14 \\
3 & 330.0 & -1.0 &  & 83.77159336 & -5.434358914 & 13.4775 & -0.07256 & 0.02798 & 0.0222 & 16.49632 & -0.79747 & 0.03127 & 0.04469 & 0.151852 & 0.017179 & 0.001491 & 0.000305 & 354.69 & 0.64 & 233.43 \\
4 & 333.0 & -1.0 &  & 83.78028406 & -5.436474308 & 15.3643 & 0.10603 & 0.01405 & 0.01473 & 18.24483 & -0.16565 & 0.03714 & 0.06303 & 0.183497 & 0.020358 & 0.001036 & 0.000866 & 9.79 & 0.19 & 217.7 \\
5 & 351.0 & -1.0 & JW 638 & 83.83343099 & -5.486611088 & 13.4015 & 0.16228 & 0.17164 & 0.02449 & 14.99455 & -0.26626 & 0.04088 & 0.05973 & 0.666812 & 0.115873 & 0.075648 & 0.003052 & 337.9 & 0.47 & 353.03 \\
6 & 355.0 & -1.0 &  & 83.84672012 & -5.476080805 & 13.6243 & -0.17387 & 0.001 & 0.00336 & 19.94556 & -0.54334 & 0.1797 & 0.2833 & 0.132205 & 0.004614 & 3.5e-05 & 0.000223 & 191.76 & 0.77 & 327.13 \\
7 & 588.0 & -1.0 &  & 83.89252874 & -5.455090951 & 11.8697 & 0.02297 & 0.01133 & 0.03316 & 15.77574 & -0.51272 & 0.02497 & 0.04258 & 0.519625 & 0.027329 & 0.003795 & 0.000757 & 134.19 & 0.44 & 355.41 \\
8 & 703.0 & -1.0 &  & 83.61902968 & -5.509038831 & 14.4798 & -0.12672 & 0.01402 & 0.0054 & 20.09559 & -0.55557 & 0.09975 & 0.28998 & 0.064618 & 0.004378 & 0.000643 & 0.000124 & 263.09 & 0.48 & 837.1 \\
9 & 943.0 & -1.0 &  & 83.82263551 & -5.537828912 & 15.0015 & -0.00517 & 0.00923 & 0.00788 & 16.19856 & -0.36395 & 0.0304 & 0.05199 & 0.138769 & 0.048765 & 0.000325 & 0.000543 & 8.86 & 0.19 & 533.54 \\
10 & 1021.0 & -1.0 &  & 83.88307815 & -5.52992167 & 14.6918 & -0.21926 & 0.01524 & 0.00222 & 19.26116 & -0.3466 & 0.0438 & 0.08264 & 0.047936 & 0.005304 & 0.000267 & 5.2e-05 & 52.26 & 0.27 & 555.71 \\
11 & 1080.0 & -1.0 &  & 83.95476263 & -5.559047659 & 16.8017 & -0.0107 & 0.01704 & 0.01693 & 19.66101 & -0.37798 & 0.04246 & 0.07579 & 0.014534 & 0.005086 & 0.000129 & 5e-05 & 183.71 & 0.22 & 782.38 \\
12 & 1161.0 & -1.0 &  & 83.89670965 & -5.448594714 & 12.1409 & 0.0594 & 0.00886 & 0.01629 & 14.04119 & 0.12296 & 0.03058 & 0.05044 & 0.495625 & 0.088725 & 0.002556 & 0.001455 & 276.57 & 0.2 & 352.27 \\
13 & 1251.0 & -1.0 & JW 709 & 83.84242884 & -5.443706152 & 11.295 & 0.19391 & 0.00395 & 0.00074 & 12.32212 & 0.30177 & 0.06507 & 0.13168 & 1.237085 & 0.540864 & 0.002806 & 0.030071 & 283.37 & 0.21 & 212.6 \\
14 & 1331.0 & -1.0 &  & 83.87796088 & -5.408648543 & 12.6713 & -0.1118 & 0.03893 & 0.02001 & 14.95313 & -0.5236 & 0.0352 & 0.05679 & 0.286316 & 0.042394 & 0.004905 & 0.000628 & 302.46 & 0.38 & 224.4 \\
15 & 1407.0 & -1.0 & JW 570 & 83.82474913 & -5.426100713 & 12.2306 & 0.04893 & 0.03399 & 9e-05 & 14.11821 & -0.0027 & 0.05342 & 0.08013 & 0.406357 & 0.076947 & 0.009162 & 0.003184 & 348.93 & 0.26 & 132.99 \\
16 & 1541.0 & -1.0 & JW 152 & 83.72936023 & -5.424836418 & 12.1775 & 0.04272 & 0.00146 & 0.00087 & 14.31968 & -0.24743 & 0.04968 & 0.07978 & 0.472033 & 0.073861 & 0.000417 & 0.001101 & 51.6 & 0.28 & 345.24 \\
17 & 1543.0 & -1.0 & JW 151 & 83.72843059 & -5.420142975 & 12.872 & -0.1633 & 0.04357 & 0.00869 & 14.36475 & 0.15954 & 0.05711 & 0.08939 & 0.235225 & 0.064372 & 0.003889 & 0.00266 & 325.05 & 0.24 & 342.58 \\
18 & 2007.0 & -1.0 &  & 83.76287689 & -5.37716716 & 11.7742 & -0.00565 & 0.00886 & 0.01562 & 14.66854 & 0.06643 & 0.03497 & 0.05352 & 0.54935 & 0.047472 & 0.002964 & 0.000624 & 192.6 & 0.2 & 205.54 \\
19 & 2396.0 & -1.0 &  & 83.8540066 & -5.400418165 & 12.8879 & 0.2529 & 0.02858 & 0.02178 & 14.59045 & -0.08418 & 0.03732 & 0.06801 & 0.888352 & 0.141468 & 0.016308 & 0.003351 & 255.61 & 0.29 & 133.26 \\
20 & 2404.0 & -1.0 &  & 83.85195641 & -5.400290404 & 13.1041 & -0.08161 & 0.05475 & 0.01808 & 14.50296 & -0.34182 & 0.03761 & 0.06597 & 0.222372 & 0.066542 & 0.004893 & 0.001752 & 293.64 & 0.28 & 126.07 \\
21 & 2464.0 & -1.0 &  & 83.82696282 & -5.401934612 & 11.734 & 0.17088 & 0.01441 & 0.00394 & 16.60982 & -0.55286 & 0.04746 & 0.08143 & 0.664391 & 0.015354 & 0.006383 & 0.000452 & 218.71 & 0.59 & 53.45 \\
22 & 2492.0 & -1.0 & [H97b] 9239 & 83.82851021 & -5.373045614 & 13.4312 & 0.09716 & 0.07249 & 0.00778 & 14.3886 & -0.3652 & 0.03982 & 0.07265 & 0.386487 & 0.117198 & 0.019113 & 0.002972 & 321.23 & 0.25 & 69.73 \\
23 & 2571.0 & -1.0 & JW 748 & 83.85038408 & -5.359054015 & 12.4144 & 0.05594 & 0.16081 & 0.02832 & 13.75559 & -0.08395 & 0.03971 & 0.05803 & 0.502218 & 0.112976 & 0.051514 & 0.002964 & 279.68 & 0.39 & 158.94 \\
24 & 2987.0 & -1.0 & JW 391 & 83.80338979 & -5.345419412 & 11.3224 & 0.17154 & 0.08285 & 0.03132 & 13.0126 & 0.00255 & 0.03685 & 0.06362 & 0.910202 & 0.146558 & 0.045441 & 0.003309 & 84.52 & 0.33 & 168.46 \\
25 & 3162.0 & -1.0 &  & 83.94418968 & -5.373398132 & 12.3236 & -0.01377 & 0.01082 & 0.00836 & 13.83882 & 0.10386 & 0.04147 & 0.06897 & 0.358728 & 0.086737 & 0.002629 & 0.001984 & 283.55 & 0.17 & 455.97 \\
26 & 3163.0 & -1.0 &  & 83.9590678 & -5.354670298 & 12.9224 & 0.0274 & 0.00393 & 0.00013 & 15.56152 & -0.21944 & 0.04992 & 0.08604 & 0.279615 & 0.035214 & 0.000489 & 0.000885 & 263.01 & 0.32 & 521.21 \\
27 & 3190.0 & -1.0 &  & 83.91659751 & -5.37407979 & 14.8903 & 0.19968 & 0.00154 & 0.00207 & 16.45456 & -0.23877 & 0.04946 & 0.08087 & 0.147277 & 0.043943 & 5.4e-05 & 0.000883 & 276.93 & 0.18 & 357.29 \\
\enddata
\tablecomments{\textit{Table \ref{Tab:CBC_klip} is published in its entirety in the machine-readable format. A portion is shown here for guidance regarding its form and content.}}
\end{deluxetable*}
\end{longrotatetable}

\begin{longrotatetable}
\movetabledown=15mm
\begin{deluxetable*}{chhhccccccccccccccccc}
\tabletypesize{\footnotesize}
\setlength{\tabcolsep}{2pt}
\tablewidth{0pt}
\tablecaption{\textit{Candidate Binaries catalog (Paper~I)}
\label{Tab:CBC_close}}
\tablehead{\colhead{ID} & \nocolhead{WFC3ID$_p$} & \nocolhead{WFC3ID$_c$} & \nocolhead{SimbadName} & \colhead{Ra$_p$} & \colhead{Dec$_p$} & \colhead{mag130$_p$} & \colhead{color$_p$} & \colhead{dmag130$_p$} & \colhead{dcolor$_p$} & \colhead{mag130$_c$} & \colhead{color$_c$} & \colhead{dmag130$_c$} & \colhead{dcolor$_c$} & \colhead{mass$_p$} & \colhead{mass$_c$} & \colhead{emass$_p$} & \colhead{emass$_c$} & \colhead{PA} & \colhead{Sep} & \colhead{SepOriC} \\ 
\colhead{(--)} & \nocolhead{(--)} & \nocolhead{(--)} & \nocolhead{(--)} & \colhead{(deg)} & \colhead{(deg)} & \colhead{(mag)} & \colhead{(mag)} & \colhead{(mag)} & \colhead{(mag)} & \colhead{(mag)} & \colhead{(mag)} & \colhead{(mag)} & \colhead{(mag)} & \colhead{(solMass)} & \colhead{(solMass)} & \colhead{(solMass)} & \colhead{(solMass)} & \colhead{(deg)} & \colhead{(arcsec)} & \colhead{(arcsec)} } 
\startdata
39 & 56.0 & 57.0 & JW  52 & 83.67010811 & -5.469291606 & 12.4762 & -0.1028 & 0.0105 & 0.02419 & 12.6109 & -0.0753 & 0.17544 & 0.01007 & 0.31939 & 0.294837 & 0.002017 & 0.026015 & 52.49 & 0.23 & 606.5 \\
40 & 238.0 & 245.0 &  & 83.7861492 & -5.483746113 & 12.5467 & 0.25674 & 0.0203 & 0.00317 & 13.2935 & 0.2766 & 0.03924 & 0.01055 & 0.304773 & 0.148041 & 0.003707 & 0.001391 & 100.11 & 1.73 & 358.21 \\
41 & 292.0 & 295.0 & JW 222 & 83.75904395 & -5.486073354 & 12.2413 & 0.13568 & 0.00245 & 0.00819 & 12.9594 & 0.1272 & 0.03694 & 0.02989 & 0.472413 & 0.288703 & 0.000702 & 0.004652 & 91.1 & 1.1 & 407.88 \\
42 & 293.0 & 297.0 & JW 235 & 83.7648425 & -5.490518796 & 12.4288 & 0.08384 & 0.017 & 0.01154 & 12.8926 & 0.20911 & 0.33596 & 0.04195 & 0.392846 & 0.291612 & 0.004566 & 0.051289 & 161.56 & 0.57 & 411.36 \\
43 & 412.0 & 409.0 & JW 422 & 83.8069495 & -5.479491472 & 12.0332 & 0.09606 & 0.23364 & 0.06372 & 12.5605 & 0.10666 & 0.20579 & 0.03443 & 0.746202 & 0.450407 & 0.136334 & 0.060834 & 70.46 & 0.37 & 326.03 \\
44 & 753.0 & 754.0 & JW  61 & 83.67804089 & -5.477043065 & 11.8368 & -0.01121 & 0.01237 & 0.0096 & 12.9269 & 0.09626 & 0.70721 & 0.06328 & 0.970304 & 0.420492 & 0.00626 & 0.325515 & 147.16 & 0.51 & 595.74 \\
45 & 803.0 & 804.0 &  & 83.71943096 & -5.495858274 & 12.1265 & 0.09143 & 0.13229 & 0.01538 & 12.7255 & 0.01296 & 0.01626 & 0.02557 & 0.620919 & 0.373322 & 0.054193 & 0.00396 & 6.01 & 0.39 & 523.0 \\
46 & 857.0 & 856.0 &  & 83.78686627 & -5.530307995 & 11.5871 & 0.17012 & 0.19642 & 0.0366 & 11.7648 & 0.11608 & 0.67489 & 0.06516 & 0.933022 & 0.847697 & 0.13524 & 0.461341 & 37.83 & 0.29 & 518.98 \\
47 & 892.0 & 897.0 &  & 83.85699037 & -5.505832787 & 11.2803 & 0.1473 & 0.00951 & 0.02309 & 12.6551 & 0.06466 & 0.0217 & 0.0111 & 0.828948 & 0.288616 & 0.004103 & 0.002726 & 316.07 & 1.67 & 440.42 \\
48 & 894.0 & 900.0 & JW 727 & 83.84501338 & -5.527000855 & 12.2636 & 0.15405 & 0.01354 & 0.01051 & 14.4277 & 0.0801 & 0.063 & 0.00569 & 0.457653 & 0.073295 & 0.003914 & 0.001426 & 291.03 & 0.79 & 503.44 \\
49 & 1248.0 & 1283.0 &  & 83.85831779 & -5.429934619 & 11.3567 & 0.25637 & 0.0281 & 0.01127 & 14.955 & 0.0025 & 0.02352 & 0.00545 & 0.798198 & 0.042626 & 0.012866 & 0.000412 & 285.94 & 1.03 & 203.63 \\
50 & 1406.0 & 1413.0 & JW 509 & 83.81740099 & -5.415657407 & 12.3663 & 0.144 & 0.01447 & 0.02249 & 12.9353 & 0.07994 & 0.01686 & 0.02084 & 0.344818 & 0.218099 & 0.003521 & 0.001498 & 41.43 & 0.54 & 93.62 \\
51 & 1415.0 & 1425.0 & [H97b] 9224 & 83.82657047 & -5.407414011 & 12.9546 & 0.03889 & 0.02408 & 0.03683 & 13.3095 & -0.0916 & 0.08866 & 0.03445 & 0.212993 & 0.146618 & 0.002143 & 0.003701 & 254.0 & 0.42 & 70.03 \\
52 & 1429.0 & 1441.0 &  & 83.81419875 & -5.433193235 & 13.5574 & -0.18268 & 0.01932 & 0.00492 & 17.1937 & -0.31123 & 0.03401 & 0.03388 & 0.128238 & 0.010471 & 0.000738 & 0.000182 & 91.71 & 1.57 & 157.45 \\
53 & 1472.0 & 1473.0 &  & 83.81199692 & -5.403242771 & 12.2095 & 0.03668 & 0.01502 & 0.00801 & 12.254 & 0.06378 & 0.00146 & 0.00032 & 0.439431 & 0.42074 & 0.004345 & 0.000387 & 336.22 & 1.34 & 54.29 \\
54 & 1483.0 & 1494.0 &  & 83.81573332 & -5.406840395 & 13.1448 & 0.08197 & 0.11398 & 0.00834 & 14.1365 & -0.1525 & 0.21425 & 0.00495 & 0.172171 & 0.073254 & 0.0083 & 0.008503 & 84.78 & 0.25 & 62.63 \\
55 & 1493.0 & 1500.0 &  & 83.80571344 & -5.398079905 & 13.7532 & -0.40305 & 0.00883 & 0.03132 & 14.3648 & -0.36328 & 0.00074 & 0.00117 & 0.10579 & 0.064912 & 0.000391 & 3.3e-05 & 280.12 & 0.98 & 55.33 \\
56 & 1527.0 & 1528.0 & JW 392 & 83.80291343 & -5.452961783 & 12.5712 & -0.0486 & 0.00591 & 0.0051 & 12.5777 & -0.05039 & 0.01511 & 0.00762 & 0.300361 & 0.299461 & 0.000737 & 0.001895 & 21.52 & 0.27 & 234.7 \\
57 & 1542.0 & 1540.0 & JW 127 & 83.71669487 & -5.411941851 & 12.6526 & 0.07137 & 0.13295 & 0.02342 & 12.7138 & 0.16735 & 0.14902 & 0.02458 & 0.361068 & 0.346167 & 0.03348 & 0.034895 & 254.17 & 0.5 & 375.45 \\
58 & 1584.0 & 1588.0 & JW 248 & 83.76820899 & -5.387204813 & 11.3141 & 0.18785 & 0.07296 & 0.02795 & 12.2152 & -0.18174 & 0.09618 & 0.00477 & 0.828116 & 0.389445 & 0.033476 & 0.025407 & 318.89 & 1.03 & 181.57 \\
59 & 1586.0 & 1587.0 & JW 201 & 83.75430427 & -5.402825758 & 11.9109 & 0.066 & 0.00149 & 0.00187 & 12.209 & 0.10544 & 0.00297 & 0.0015 & 0.483427 & 0.382948 & 0.000425 & 0.00079 & 247.41 & 1.05 & 236.2 \\
60 & 1590.0 & 1593.0 & JW 190 & 83.74711256 & -5.392456476 & 13.2541 & -0.03134 & 0.01629 & 0.00829 & 13.8071 & -0.03113 & 0.04928 & 0.03227 & 0.151572 & 0.100371 & 0.000789 & 0.002337 & 350.7 & 0.6 & 257.49 \\
61 & 1595.0 & 1600.0 &  & 83.75085224 & -5.402563056 & 14.8182 & 0.1399 & 0.01151 & 0.004 & 16.5954 & -0.322 & 0.01252 & 0.01322 & 0.070804 & 0.019725 & 0.000326 & 0.000119 & 10.76 & 1.49 & 248.2 \\
62 & 1617.0 & 1621.0 & JW  81 & 83.69329687 & -5.408822911 & 11.6392 & 0.12523 & 0.01411 & 0.01518 & 13.1086 & 0.04691 & 0.01475 & 0.01212 & 0.68968 & 0.199359 & 0.006249 & 0.001087 & 274.14 & 1.48 & 456.27 \\
63 & 1986.0 & 1987.0 & JW  63 & 83.6789671 & -5.335279829 & 11.0598 & 0.12879 & 0.01307 & 0.02797 & 11.164 & 0.07703 & 0.11576 & 0.01705 & 1.085792 & 1.043077 & 0.007172 & 0.060041 & 351.92 & 0.27 & 539.42 \\
64 & 2020.0 & 2021.0 &  & 83.73318716 & -5.36862182 & 14.1939 & -0.04466 & 0.02008 & 0.00515 & 14.3152 & -0.0748 & 0.10923 & 0.06099 & 0.071881 & 0.067254 & 0.000491 & 0.004651 & 329.48 & 0.28 & 316.63 \\
65 & 2039.0 & 2040.0 & JW 128 & 83.71736831 & -5.375495097 & 11.0921 & 0.14235 & 0.26149 & 0.00092 & 11.4305 & 0.0573 & 0.27645 & 0.02362 & 0.917212 & 0.766359 & 0.214105 & 0.197388 & 213.64 & 0.41 & 367.93 \\
66 & 2042.0 & 2043.0 & JW 124 & 83.71579716 & -5.360911408 & 13.1125 & 0.00998 & 0.00071 & 0.01474 & 13.1251 & -0.00467 & 0.14355 & 0.01523 & 0.218997 & 0.215652 & 6.2e-05 & 0.012077 & 12.53 & 0.44 & 384.25 \\
\enddata
\tablecomments{\textit{Table \ref{Tab:CBC_close} is published in its entirety in the machine-readable format. A portion is shown here for guidance regarding its form and content.}}
\end{deluxetable*}
\end{longrotatetable}

\begin{figure}[t!]
\begin{center}
\includegraphics[width=0.49\textwidth]{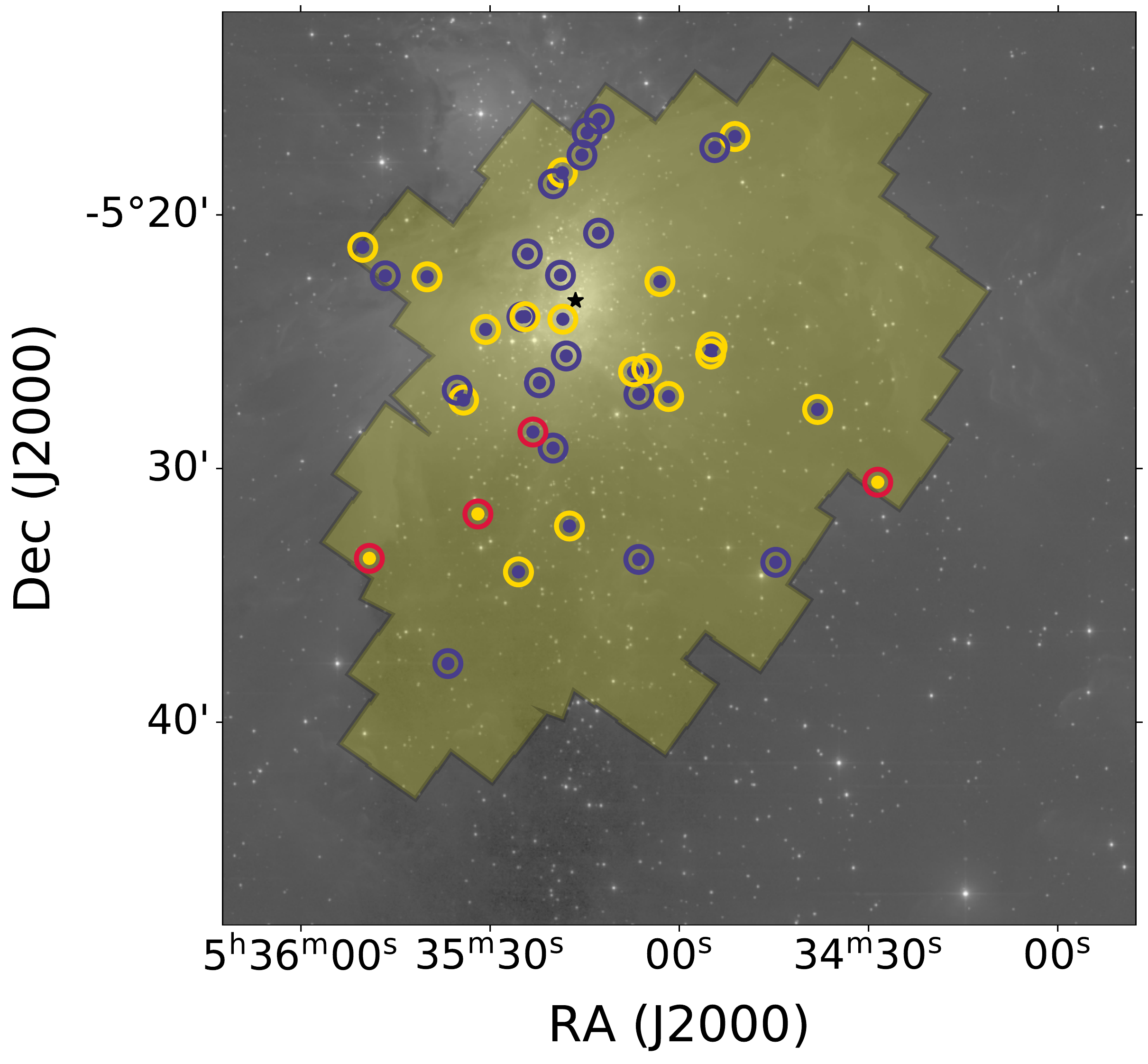}
\caption{Large scale view of ONC. The shaded yellow area indicates the field covered by the WFC3 observations, overlaid on the 2MASS J-band image of the region (in grayscale). The black star marks the position of $\theta^1$Ori-C. Colored open circles and dots mark the positions of new candidate binary systems, where the dots refer to primary stars and the open circles refer to candidate companions. The colors encode the mass of the object: blue = stellar mass object, yellow = brown dwarf, red = planetary mass object.\label{Fig:area}}
\end{center}
\end{figure}

\begin{figure}[h!]
\begin{center}
\includegraphics[width=0.49\textwidth]{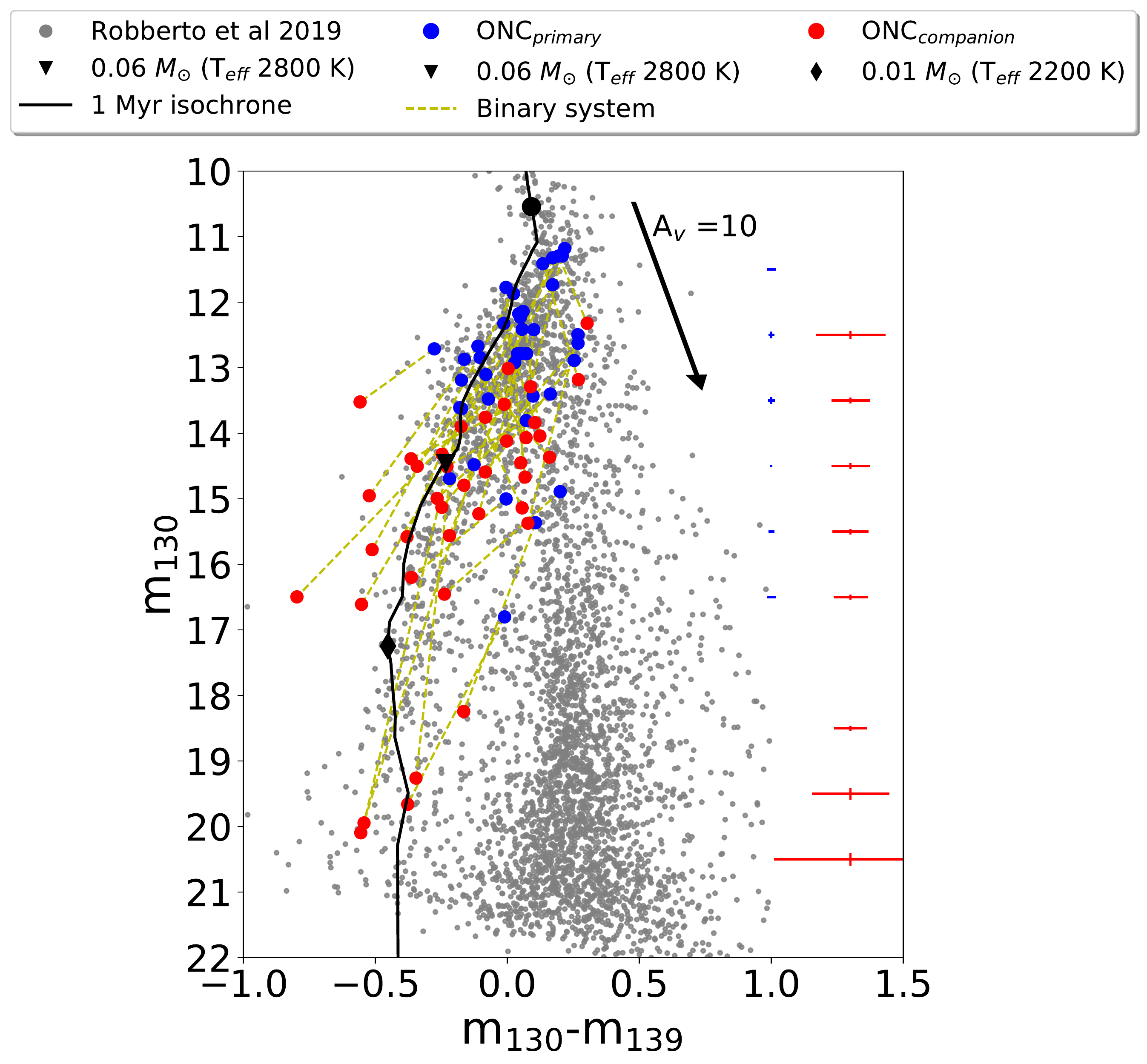}
\caption{Recreation of the CMD from Figure \ref{Fig:plain_cmd}, now including the candidate binary systems.
The black line is the 1 Myr isochrone of Paper~I, with three characteristic masses marked ($M_{\sun} = 0.02,0.08$ and $1$).
The blue (red) crosses to the right show the average uncertainties for the primaries (companions) in each magnitude bin. The yellow dotted lines join the components of each candidate  binary system. \label{Fig:cmd}}
\end{center}
\end{figure}

In Figure \ref{Fig:area} we show the position of each candidate cluster binary projected against the survey area of the WFC3 survey while
Figure \ref{Fig:cmd} shows the color magnitude diagram for the entire region with the locus of the KLIP candidate cluster binaries.


\subsection{Wide binaries}\label{subsection:ONC candidate binary catalog}
Previously, binary systems that were well resolved in Catalog~I were excluded from our analysis, which was designed to discern close companions hidden under the PSF wings of apparently single stars. 
We now expand our close companion catalog by adding the wider pairs from Catalog~I -- 58 systems with projected separation $d<1.8''$ (choosing as limit the maximum distance at which we still measure an increase in the number density of stars -- see Figure \ref{Fig:VBinaries} in  Section \ref{section:Crowding and apparent pairs}) and colors compatible with cluster membership for both sources. 
The brightest star of each pair is generally taken as the primary.
Adopting the F130N filter photometry in Catalog~I to estimate their masses,
we obtain for the primaries values in the range M$_P = 0.02-1.08\: M_{\sun}$, while for the companions we find M$_C= 0.01 -1.04 \: M_{\sun}$. Their photometry and resulting physical parameters are listed in Table \ref{Tab:CBC_close}, and a gallery of images is shown in Figures \ref{Fig:Binary_gallery_close} in Appendix \ref{app_section:Gallery of binaries} (similarly to what we presented for the KLIP pairs). 

It should be noted that we did not attempt to find faint substellar companions under the PSF wings of the Paper~I binaries, as this goes beyond the current capabilities of our implementation of the KLIP algorithm. Our search strategy, therefore, is generally biased against finding triplets or higher order systems.


\subsection{Master Catalog}
Hereafter we will refer to the combination of Table~ \ref{Tab:CBC_klip} and Table~\ref{Tab:CBC_close} as to our Master Catalog. 
The Master Catalog contains 97 pairs of stars with separations between $0.16''-1.73''$ (corresponding to $66-697$~ AU) and masses in the range M$_P=0.015 - 1.27$ M$_{\sun}$ and M$_C=0.004 - 1.04$ M$_{\sun}$ for the primary and companion, respectively. 

\begin{figure}[t!]
\begin{center}
\includegraphics[width=0.49\textwidth]{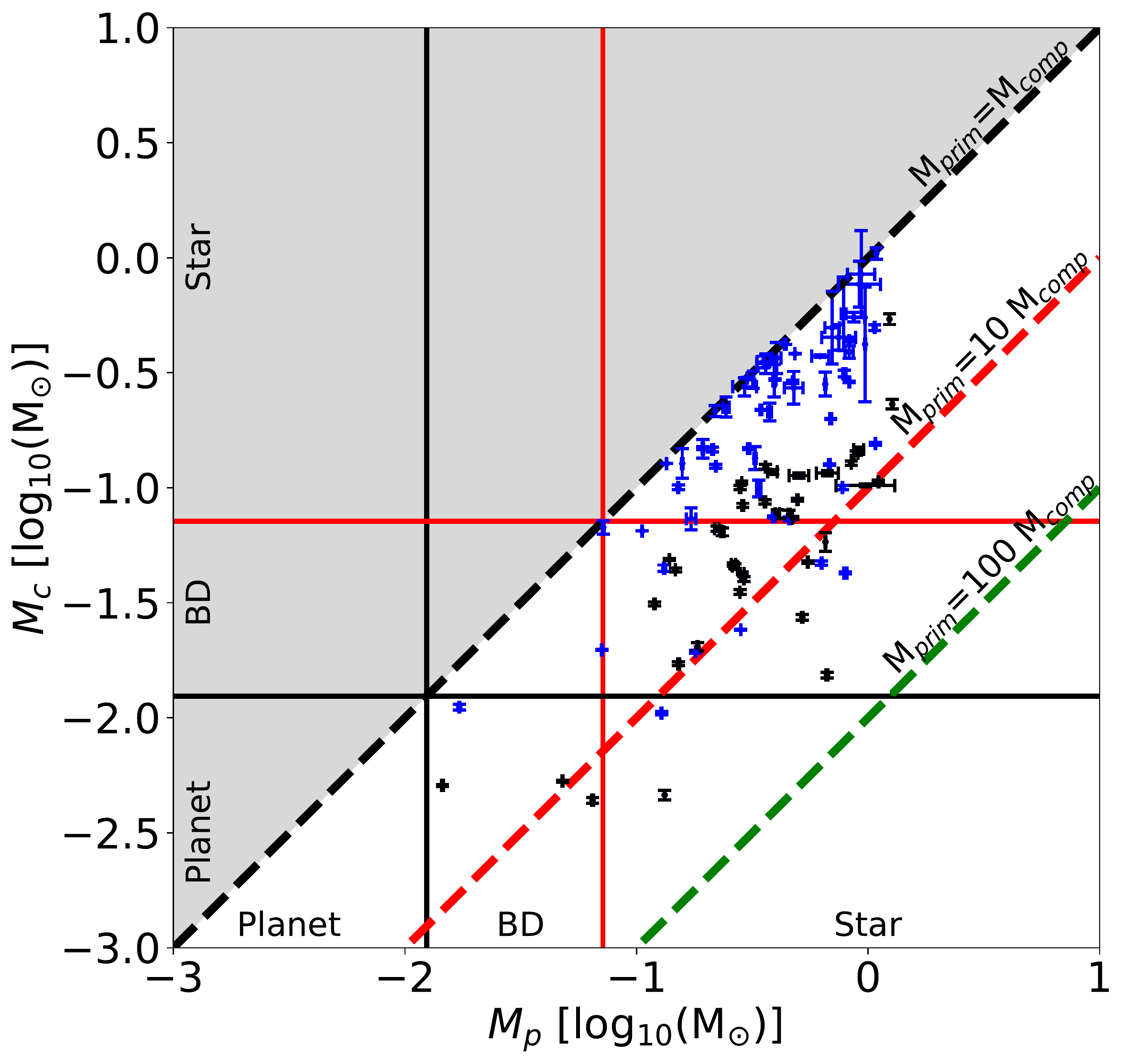}
\caption{The plot shows the relation between mass of the companion and the primary for each candidate binary system, with blue indicating pairs found in Catalog~I and black indicating pairs found in this work. The three dotted lines mark  where $M_{P}=M_{C}$ (black dotted line), $M_{P}=10\,M_{C}$ (red doted line), $M_{P}=100\,M_{C}$ (green dotted line). The planetary mass objects are separated from the brown dwarf mass objects with a black solid line while the brown dwarf mass objects and stellar mass objects are divided by a red solid line \label{Fig:ONCONC_All}}
\end{center}
\end{figure}

Figure \ref{Fig:ONCONC_All} shows the relation between primary and companion masses for all sources in the Master Catalog, with relative error bars and colors identifying the KLIP binaries (black) vs. Catalog~I binaries (blue). The diagonal lines mark the loci of systems with primary mass equal to 1, 10 and 100 times the mass of the companion, 
whereas the horizontal and vertical lines indicate the boundaries between stellar, brown dwarfs and planetary mass objects. 
The number of systems in the areas delimited by these lines is given in Table \ref{Tab:MC det}. Overall we observe a primary star-to-brown-dwarf ratio (SBdR) N(0.1-1.27 $M_{\sun}$)/N(0.014-0.07 $M_{\sun}$) $=15.16$, while the same ratio for isolated stars in the ONC ($\sim 3.8$ evaluated from Catalog~I or $3.3^{+0.8}_{-0.7}$ from \citealp{Slesnick2004,Andersen2008}) and in the field (5.2 or 6 from \citealp{Bihain2016} and \citealp{Kirkpatrick2012} respectively) is much smaller. Because the two SBdRs are different from each other (binaries vs singles in ONC/field), this may suggest a preference for companions to form around stellar mass primaries instead of brown dwarf in the ONC. This discrepancy may be due to the intrinsic difficulty in detecting companions around fainter primaries, so we evaluated the SBdR from our \textit{completed} catalog of binaries obtaining $\sim 10.6\pm0.3$. Even if we consider the \textit{completed} distribution of binaries, we still observe a preference for companions to form around primaries in the stellar mass regime compared to brown dwarf mass.


\begin{deluxetable}{ccccc}
\tablecaption{Summary of detections in Master Catalog.\label{Tab:MC det}}
\tablehead{\colhead{} & \colhead{}& \multicolumn{3}{c}{Primary}\\
& & \colhead{Star} & \colhead{Brown Dwarf}& \colhead{Planet}}
\decimals
\startdata
\multirow{3}{*}{Companion} & Star        & 63  & - & -\\
& Brown Dwarf & 26  & 2 & -\\
&Planet      & 2   & 4 & 0\\
\enddata
\end{deluxetable}

\subsection{Crowding and apparent pairs}
\label{section:Crowding and apparent pairs}
Given the increasing stellar density toward the inner regions of the cluster, one may expect to find apparent pairs due to chance alignments, i.e. cluster members that have small projected separation but are physically unrelated. Assuming a random distribution, one can use estimators like a two-point correlation function to evaluate the probability of observing a pair at a particular separation. Departures from random probability may indicate the presence of real close binaries.

To perform this analysis, we follow \cite{Jerabkova2019}, building the so-called Elbow plot \citep{Gladwin1999,Larson1995}, showing the number density of detected targets ($\Sigma$) as a function of the separation on-sky ($\theta$). As shown by \cite{Gladwin1999}, the presence of an elbow in this distribution graphically indicates the presence of resolved binaries. 

\begin{figure}[t!]
\begin{center}
\includegraphics[width=0.49\textwidth]{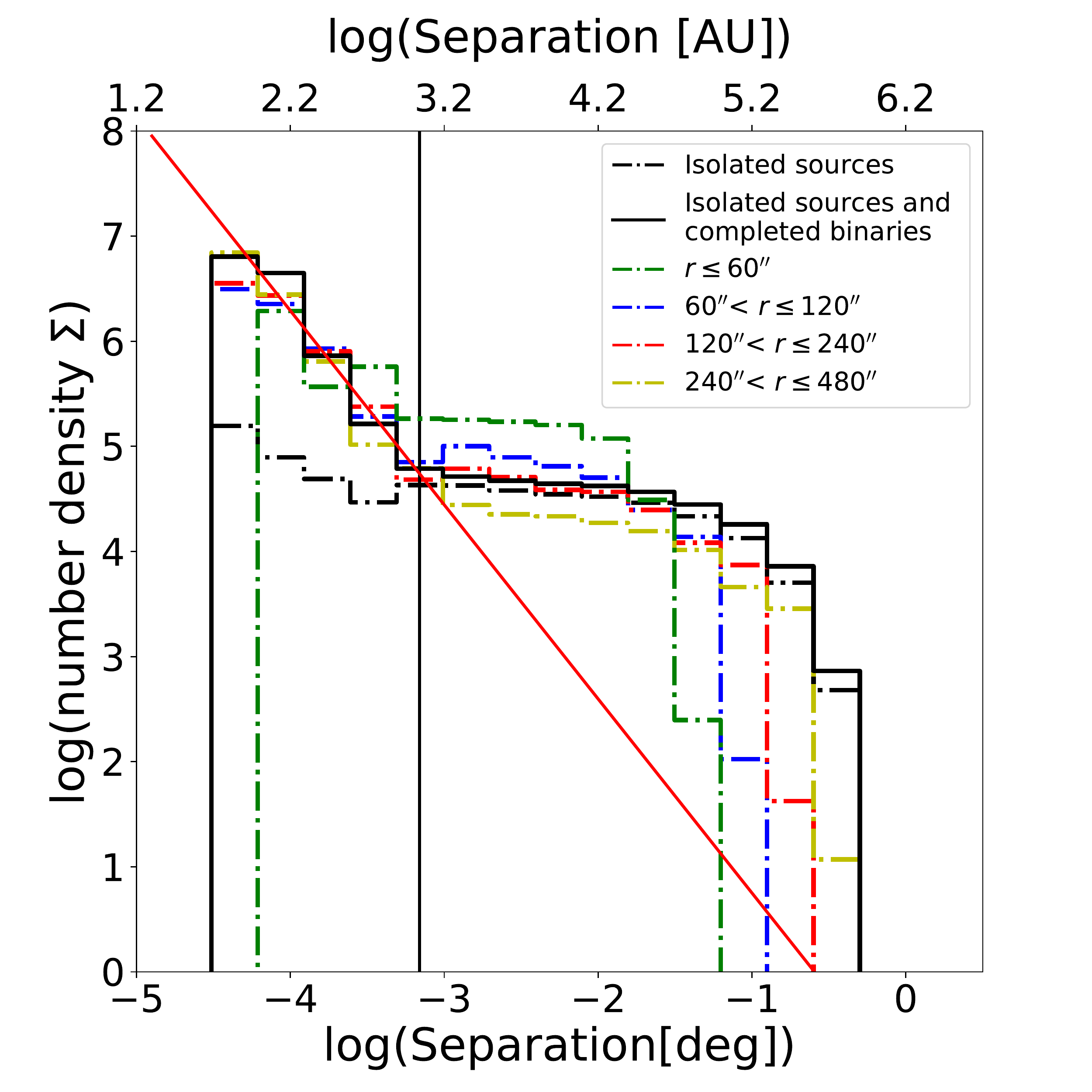}
\caption{Elbow plot showing the number density of stars $\Sigma$($\theta$)  as a function of the on-sky separation $\theta$. The dash-dotted black histogram shows data from the cluster selected isolated sources from Catalog~I (no binaries), while the solid black histogram shows the same data where we added the \textit{completed} distribution of binaries obtained from the Master Catalog (isolated sources plus binaries). The colored dash-dotted histograms shows the distribution obtained from the isolated sources plus binaries data cut at different distances ($r$, in the legend) from the position of $\theta^1$Ori-C. 
The strong gap between the elbow of the “isolated sources” and the other ones shows that our binaries have to be “bound”. The vertical black line show the transition point between the flat portion of the $\Sigma$($\theta$)  and the start of the elbow in our plot ($\sim10^3$ AU). The red line shows the fit of the elbow for the completeness-corrected Master Catalog (slope: $-1.85 \pm 0.32$). }
\label{Fig:VBinaries}
\end{center}
\end{figure}

Figure \ref{Fig:VBinaries} shows the Elbow plot derived from the cluster selected isolated sources of Catalog~I (black dash-dotted histogram) and the same data where we also add the \textit{completed} distribution of binaries obtained from the Master Catalog (black solid histogram). To investigate how the excess of binaries varies with the radial distance from the cluster center, the figure also shows the results for four different rings centered around the position of $\theta^1$Ori-C. 
Overall the different distributions agree with each other, all showing a clear overabundance of multiple systems starting at $\sim 10^3$~AU (black vertical line).  This result is in agreement with \cite{Scally1999} who suggested, based on a common proper motion study, that there should be no binaries wider than 1000 AU. Using GAIA DR2 data in combination with ground-based visible images, \cite{Jerabkova2019} finds for the ONC that the overabundance of multiple systems stars at $\sim 3000$~AU. Our data, reaching fainter objects with the diagnostic power to separate  cluster members from background sources, lend support to Scally's findings. Moreover, 
fitting the elbow part of the global $\Sigma$($\theta$) distribution we find a slope  $-1.98\pm0.30$ (red line), in excellent agreement with typical values for young clusters  \citep{Gladwin1999} as well as for early studies of the ONC in particular \citep{1998MNRAS.297.1163B}.
These results indicate that the true population of  binaries in the ONC has been reliably assessed, and that no overestimate is introduced by our completeness correction. 
\subsection{Comparison with previous HST surveys }\label{subsection :Match with Reipurth}

\begin{deluxetable}{ccccc}
\tablecaption{The table shows the number of matched binaries between our catalog and previous surveys. The columns shows the number of binaries matched to our Master Catalog (Cluster) or detected and rejected because a component was assigned to the background (Background), the number of binaries matched to a single star in Catalog~I but not present in the Master Catalog (Unresolved) or not matched at all (Not Matched) \label{Tab:Comp surv}}
\tablehead{\colhead{}& \colhead{Cluster} & \colhead{Background}& \colhead{Unresolved}& \colhead{Not Matched}}
\setlength{\tabcolsep}{2pt}
\decimals
\startdata
Reipurt et al. 2007     & 53  & 16 & 8 & 14\\
Duchene et al. 2018     & 0   & 0  & 7 & 7\\
DeFurio et al. 2019     & 3   & 3  & 5 & 3\\
\enddata
\end{deluxetable}

\cite{Reipurth2007a}, using HST/ACS H$\alpha$ images from GO-9825 with 50~mas pixel size (corresponding to about 20~AU, 2.5 times smaller than our WFC3-IR data) performed a major survey for visual binaries in the ONC probing a range of separations similar to ours. More recently, \cite{DeFurio2019} used PSF fitting to find close pairs in HST/ACS F555W (V-band) images from GO-10246 to probe separations smaller than 160~AU. These surveys, like those performed using ground-based Adaptive Optics systems, in particular \cite{Duchene2018a}, are complementary to our study as they target brighter and bluer (i.e. typically more massive) sources at smaller separations. 
Comparing the systems in our Master Catalog with those reported in the three aforementioned surveys, we obtain the results listed In Table \ref{Tab:Comp surv}. The columns list the number of targets we identify as cluster members (``Cluster''), those having at least one component classified as background source (``Background''), those appearing unresolved in our data even after KLIP processing (``Unresolved''), and those that do not match any source in our catalog (``Not matched''). If we exclude the binaries that were previously identified in \cite{Reipurth2007a} and \cite{DeFurio2019} and those identified in Paper~I, we are left with 21 new candidate binaries uncovered by the KLIP algorithm. These new candidate detection span a range of primary masses between $0.014-0.127$ M$_{\sun}$, companion masses $0.004-0.23$ M$_{\sun}$, separations $0.16-0.76''$ and completeness between $17\% - 87\%$ with $49\%$ as median value.

\begin{figure}[t!]
\begin{center}
\includegraphics[width=0.49\textwidth]{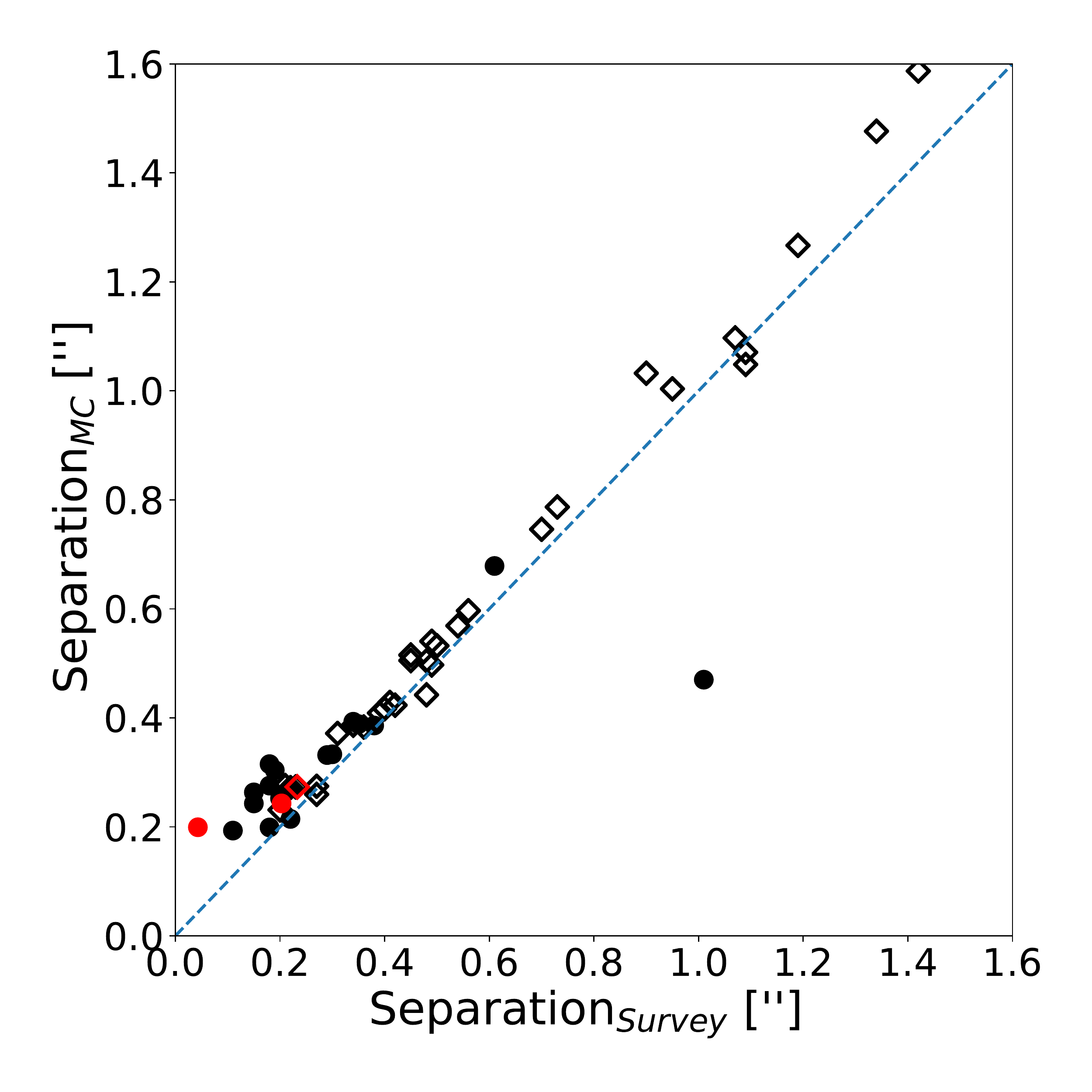}
\caption{Comparison of separation between \citeauthor{Reipurth2007a} (black) and \citeauthor{DeFurio2019} catalogs (red)  vs \emph{Master Catalog} for matched cluster-cluster binaries. The dots mark the position of the matched binaries obtain through KLIP PSF subtraction, while the hollow diamond marks the matched binaries obtained from Paper~I.
The blue dotted line show the locus of point where Separation$_{ONC}$~=~Separation$_{Survey}$} \label{Fig:CompReip_sep}
\end{center}
\end{figure}

Figure \ref{Fig:CompReip_sep} shows a comparison between the separations reported in our Master Catalog versus those given by \citeauthor{Reipurth2007a} (black) and \citeauthor{DeFurio2019} (red). Overall, there is excellent agreement between our values and those reported by these surveys,
with only one major discrepancy against the \citeauthor{Reipurth2007a} catalog: their Source JW~638 is listed as having a companion at $\sim 1''$ separation, whereas our IR images \citep[as well as the ACS visible images of][]{Robberto2013} show a closer companion at separation $\sim 0.4''$ (see Fig. \ref{Fig:KLIP_gallery0}, ID 7). If we exclude this detection, the average scatter of separations between our catalog and the others is $\sim 0.05''$, less than $~1/2$ WFC3 pixel.

\section{Discussion}
\label{section:discussion}

\subsection{Binary Frequency}\label{Section:BinaryFrequency}
The multiplicity function (MF) of multiple systems is defined as:
\begin{equation}
    MF={N_{mult}\over N_{mult}+N_{single}}
\end{equation}
where $N_{mult}$ and $N_{single}$ are the number of multiple and single star systems in the sample. In Table \ref{Tab:MFtable} we report the MF values for a) the Master Catalog (``all''); b) for the Master catalog split in two different bins of primary mass ("star  or "BD"), and c) three different primary mass bins (B0, B1 and B2) having the same number of systems in each bin. Table~\ref{Tab:MFtable} shows that the fraction of binaries among stellar mass objects is 3 times larger than among substellar mass objects, for the separation range we are considering. The deficit of very low mass binary systems  remains regardless on how the limits are defined, as shown by the bottom half of the table.
\begin{deluxetable}{cccc}
\tablecaption{Multiplicity Fraction (MF) for the complete sample (first row) and different subsamples of primary masses in separation rage of 0.16''-1.73''.\label{Tab:MFtable}}
\tablehead{\colhead{Label} & \colhead{Primary mass [M$_{\sun}$]} & \colhead{MF [$\%$] }}
\decimals
\startdata
All  & 0.01-1.27  & 11.5  $\pm$ 0.9 \\
Star & 0.08-1.27  & 14.6  $\pm$ 1.1 \\
BD   & 0.01-0.08  & 4.6   $\pm$ 1.3 \\
\hline
\hline
B0 & 0.50-1.27   & 21.6 $\pm$ 2.9 \\
B1 & 0.28-0.50   & 14.5 $\pm$ 1.9 \\
B2 & 0.01-0.28   & 6.8 $\pm$ 1.0 \\
\enddata
\end{deluxetable}

A variety of MF values have been previously reported in the literature for the ONC. \cite{Petr1998} looked for binaries in the inner $40''\times 40''$ around  the Trapezium, finding $MF=5.9\%\pm 4.0\%$ in the separation range $0.14''-0.5''$ (63-225 AU). In a similar separation range we obtain MF=$8.1\%\pm 0.8\%$. \cite{Kohler2006} performed a survey of the periphery of the ONC at 5–15 arcmin (0.65–2 pc) from the cluster center, probing separations from $0.1''-1.2''$ and primary masses from $0.1-2\:M_{\Sun}$, finding MF=$5.1\%\pm2.7\%$; for a similar range of mass and separation we find MF = $13.0\pm1.1$. \cite{Reipurth2007a} report MF=$8.8\%\pm1.1\%$ in the range of separations $0.17''-1.69''$ ($67.5 - 675$~AU)  while we find $10.8 \% \pm 0.9\%$. 
In general, we obtain larger MF values than previous ONC studies because the combination of \textit{HST}/WFC3 and KLIP allows us to unveil a larger number of faint companions at low angular separations. Still, in comparison with other star forming regions, our multiplicity function is $\sim 2$ smaller than e.g. Taurus over a similar separation range \citep{Duchene2013}. On the other hand, comparing our result with the binary frequency in the field obtained by \cite{Duquennoy1991} for a similar range of separations, we find approximately the same binary frequency between the field and the ONC. This result is also in agreement with \cite{DeFurio2019} where the author find that the low-mass star binary population of ONC is consistent with that of the Galactic Field over mass ratio $0.6 - 1$ and separation $30 -160$ AU.


\subsection{Binary Separation}
\label{subsection:Separation distribution}
\begin{figure}[t!]
\begin{center}
\includegraphics[width=0.49\textwidth]{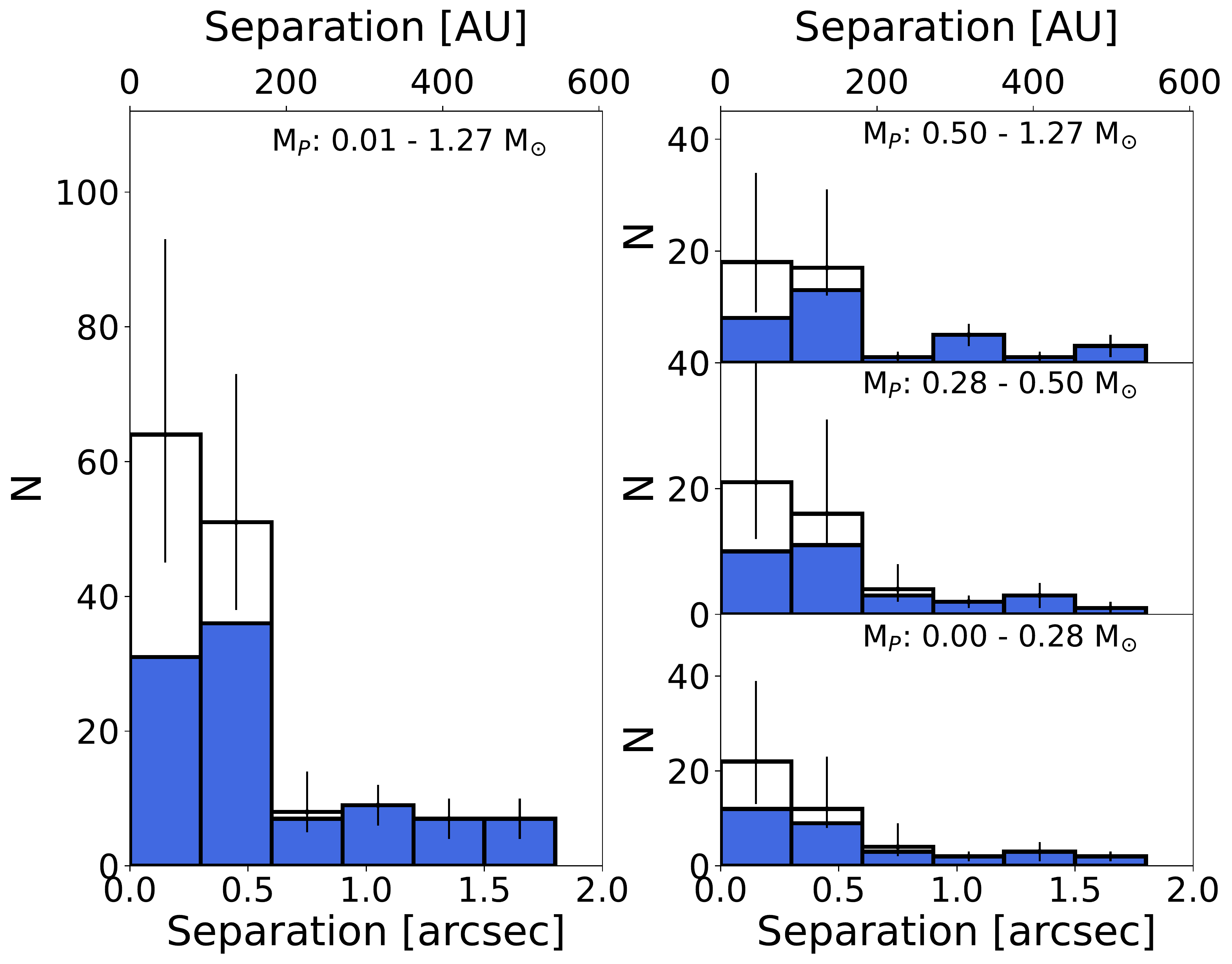}
\caption{Separation distribution for our full sample (left) and different subsamples of primary masses as in Table \ref{Tab:MFtable}, in bins of $0.3"$.
The number of observed companions in each bin is given by the blue histogram; the hollow black histogram -- as before -- indicates the completeness-corrected value. Error bars are determined according to Poisson statistics. \label{Fig:HistSep}}
\end{center}
\end{figure}

The left panel of Figure~\ref{Fig:HistSep} shows the distribution of projected separations in the Master Catalog in bins of $0.3''$ before and after completeness correction. The right panel shows  histograms of the separations for the three equally populated mass intervals B0, B1, B2 introduced in Section~\ref{Section:BinaryFrequency}. Overall, the separation distribution is peaked toward small values $ \lesssim 0.6 ''$, or 240~AU. At larger distances, the distribution shows a plateau, both results being consistent with what has been already reported by \cite{Reipurth2007a}. 

\begin{deluxetable}{ccccc}
\tablecaption{Typical timescale for free floaters for the close and wide binary populations\label{Tab:tff}}
\tablehead{\colhead{} & 
             \colhead{Primary mass} & 
                \colhead{Companion mass} & 
                    \colhead{Separation}& 
                        \colhead{$\tau_{ff}$}\\
                    &                           
             [M$_{\sun}$] &
                [M$_{\sun}$] &
                    ['']    & 
                        [Myr] }
\decimals
\startdata
close & 0.45 & 0.22 & 0.32 & 111\\
wide  & 0.36 & 0.17 & 1.13 & 37\\
\enddata
\end{deluxetable}

\cite{Spurzem2009} have analyzed the disruption of planetary systems in the ONC. Their numerical simulations indicate that moderately
close stellar encounters can cause the disruption of planetary systems. 
They find that the ejected planets have typically low velocity dispersion and in young clusters can be retained by the cluster potential and appear as free floaters. 
Table \ref{Tab:tff}, based on \cite{Spurzem2009}  Eq. 36 and Eq. 37, shows the typical timescale to get a free floater ($\tau_{ff}$) for the "close" ($0.1'' -  0.6''$) and "wide" ($0.6'' - 1.5 ''$) population of binaries assuming our typical values for the primary and companion mass, and system separation. 
Considering the total number of systems that may harbor a companion, disruptions 
can be expected, in particular for 
the wide binary population in the central region of the cluster which had statistically enough time to undergo at least one strong gravitational encounter. 
The observed spectrum of binary separations, in particular the discontinuity between close and wide binaries at 0.6'' (240~AU), can thus be attributed to stellar encounters, as anticipated by \cite{Reipurth2007a}.  

\subsection{Binary Separation vs. Distance from the Cluster center}
\label{section:SepOriC}
In this section we examine if the close and wide binaries,
separated at 240~AU, can be isolated as two distinct populations depending on the distance from the cluster core. 

\begin{figure}[t!]
\begin{center}
\includegraphics[width=0.49\textwidth]{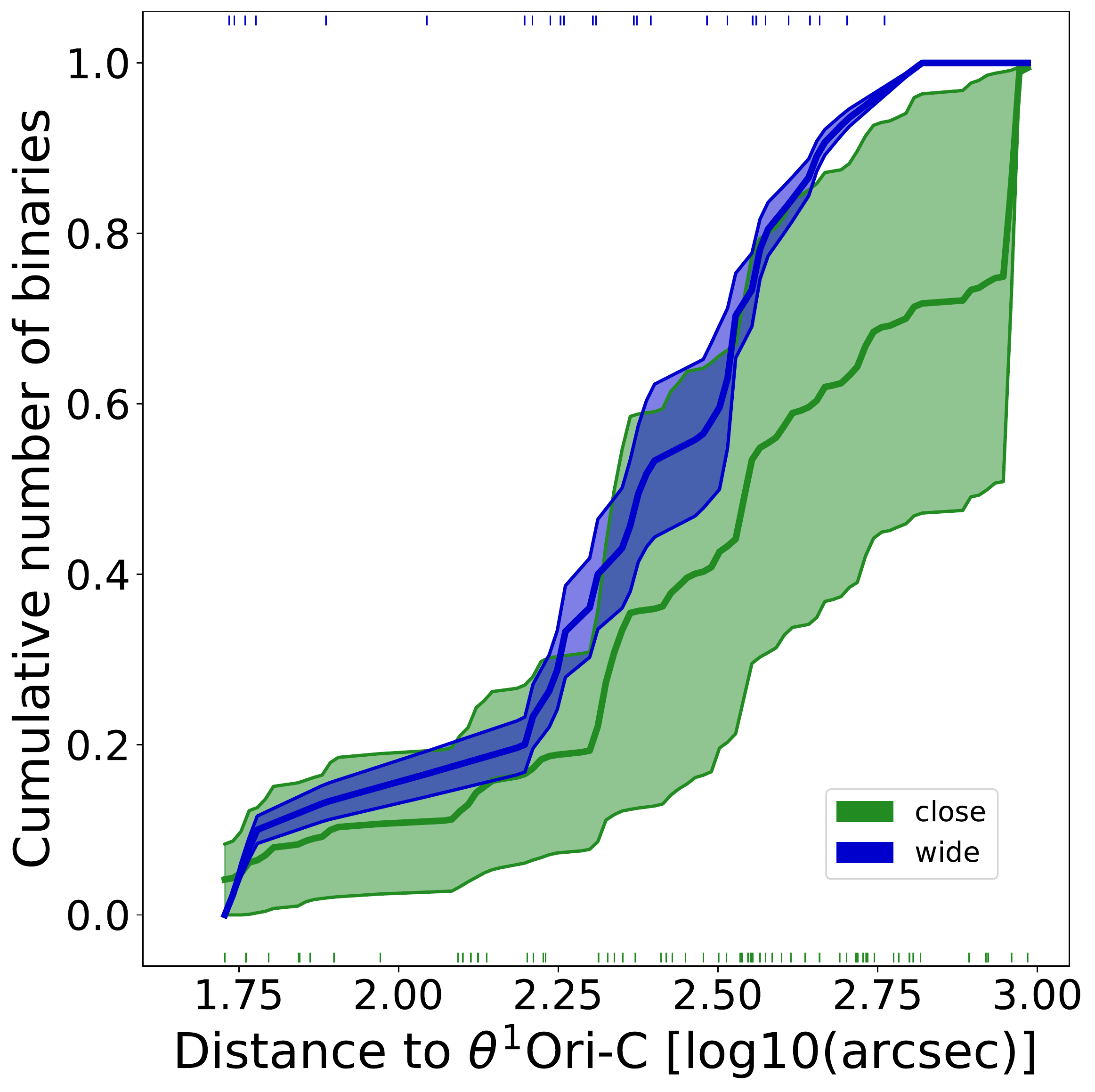}
\caption{The green (blue) area shows the family of curves obtained through the simulations of \emph{completed} (see text) cumulative distribution for close (wide) binaries as a function of the distance from $\theta^1 Ori$. The green (blue) vertical ticks at the bottom (top) of the plot show the distance from the core for each system used to generate the synthetic populations.  \label{Fig:CumDist}}
\end{center}
\end{figure}

To perform this analysis, we study the completeness-corrected cumulative distributions of close and wide binaries, but instead of simply applying a completeness correction to our observations, we estimate the "true" number of underlying objects required to observe an object given the estimated completeness $\mathcal{C}$.
The number of missed detections for each successful detection at completeness $\mathcal{C}$ is modeled as a negative binomial distributions representing the number of failures $f$ occurring before a number of successes $s$ is observed, assuming a probability $p$ of single success. We define the specific shape of the negative binomial distribution (for each detection) by using the value $p=\mathcal{C}$ for the individual trial success probability, and $s=1$.
Using this negative binomial distribution, we extract a random number of "failures", i.e. undetected companions, that were not observed due to noise and/or incompleteness. We then assign to each of these systems a distance from the center similar to that of the actually observed systems.
Finally, we iterate over the sample of observed binaries to obtain a single realization of a "complete" binary population and repeat this procedure one thousand times to obtain the \emph{completed} cumulative distributions shown in Figure \ref{Fig:CumDist} for close (green) and wide (blue) binaries. 
For each iteration we perform a 2-sample Kolmogorov-Smirnov test (KS-test) on the \emph{completed} populations of close and wide binaries as a function of the distance from the core of the cluster. 
For $\sim 48 \%$ of the KS-tests we obtain a p-value below 0.01. At this level of confidence, we can not safely reject the hypothesis that the two samples are drawn from the same distribution. This suggests that the two populations may be different with respect to their spatial distribution. 

\subsection{Mass distribution}
\label{IMF}

\begin{deluxetable}{ccccc}[!t]
\tablecaption{Fitted values for $\Gamma$ in the broken power law in the mass range $0.015-1.27\: M_{\sun}$.\label{Tab:Gammas}}
\tablehead{\colhead{Group} & \colhead{$\Gamma_1$} & \colhead{$\Gamma_2$} & \colhead{$log\:M$}}
\decimals
\startdata
Primaries            & -0.9 $\pm$ 0.5    & 0.2 $\pm$ 0.4       & -0.3 $\pm$ 0.1\\
Companions           & -0.6 $\pm$ 0.7    & 0.9 $\pm$ 0.6       & -0.8 $\pm$ 0.2\\
\enddata
\end{deluxetable}

\begin{figure}[t!]
\begin{center}
\includegraphics[width=0.49\textwidth]{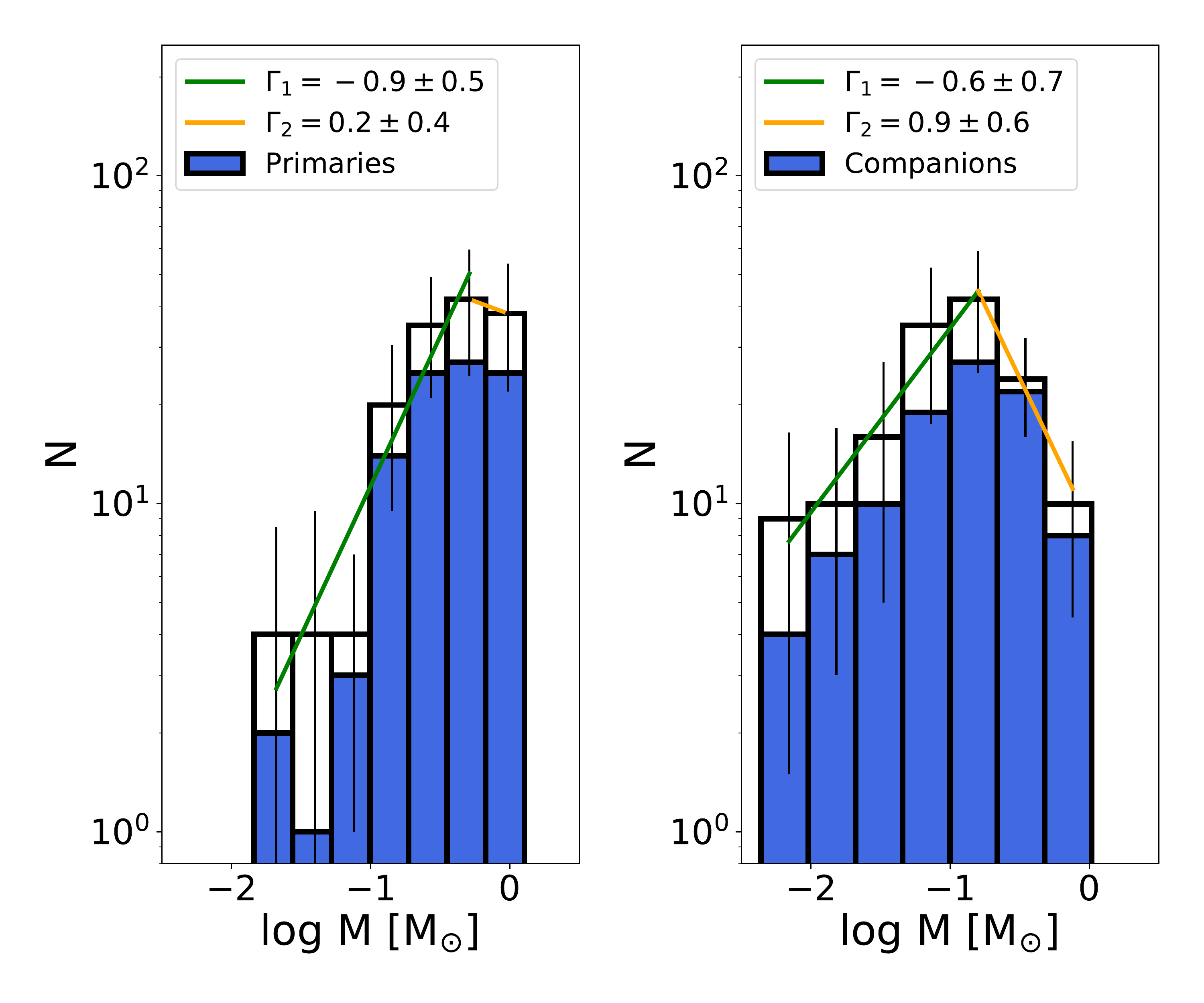}
\caption{Histogram of masses for primaries and companions. The blue histogram shows the number of companions, while the hollow histogram is corrected for completeness. To bin the two distributions we used {\sl Scott's method} \citep{Scott1979}, where the optimal histogram bin width takes into account data variability and data size by asymptotically minimizing the integrated mean square error. \label{Fig:HISTmasses}}
\end{center}
\end{figure}

\begin{figure*}[t!]
\begin{center}
\includegraphics[width=0.48\textwidth]{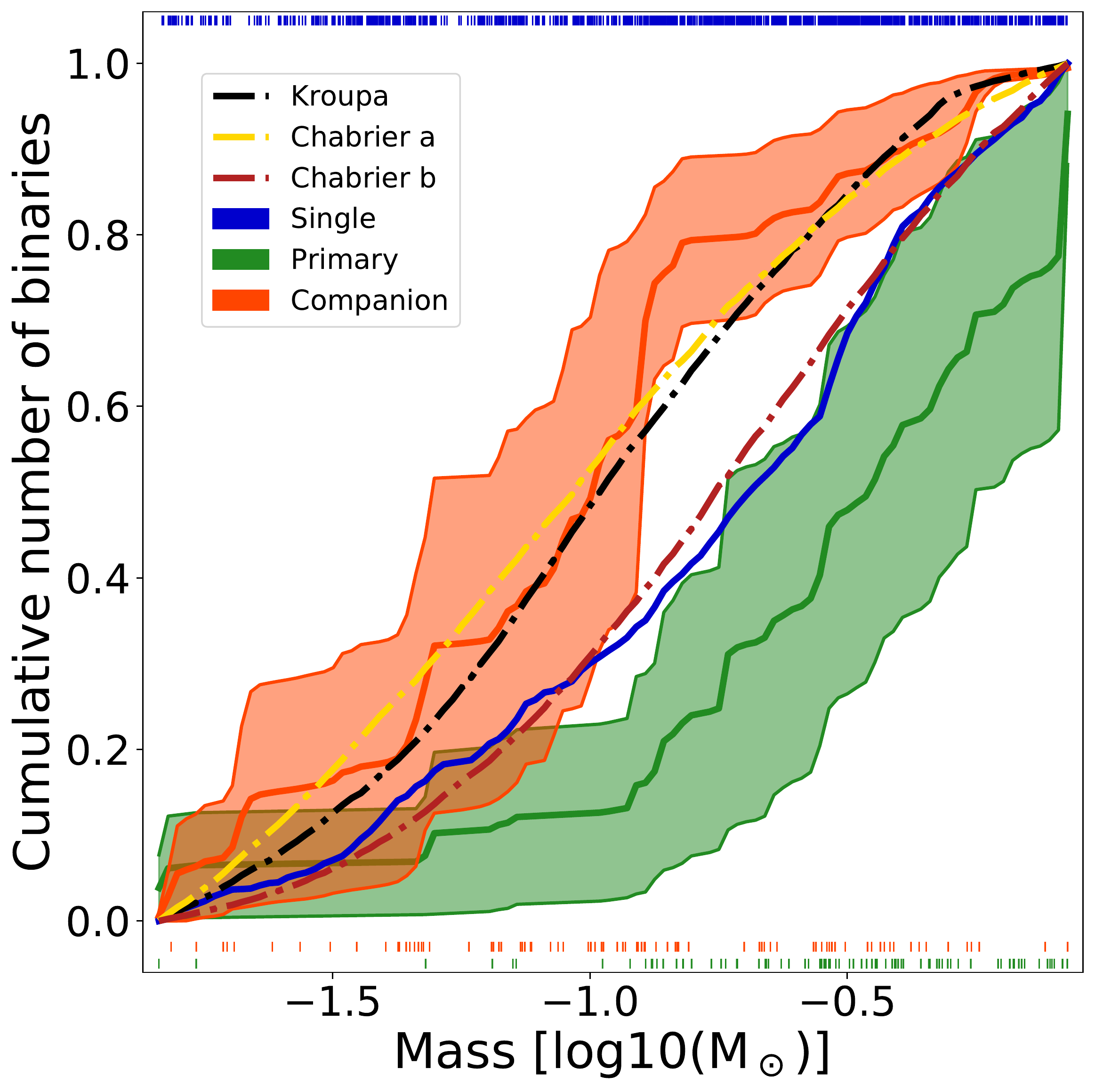}
\includegraphics[width=0.48\textwidth]{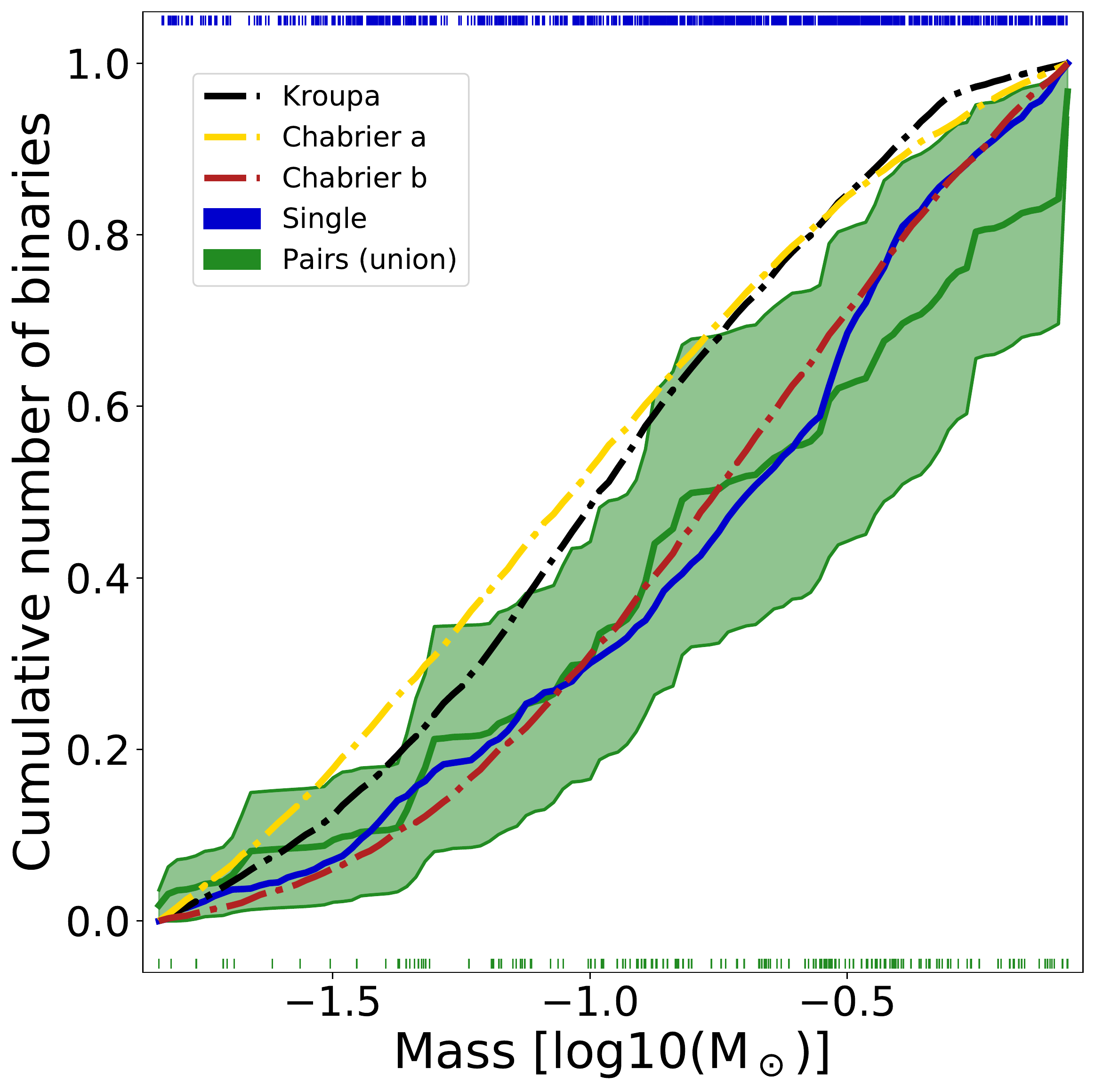}
\includegraphics[width=0.48\textwidth]{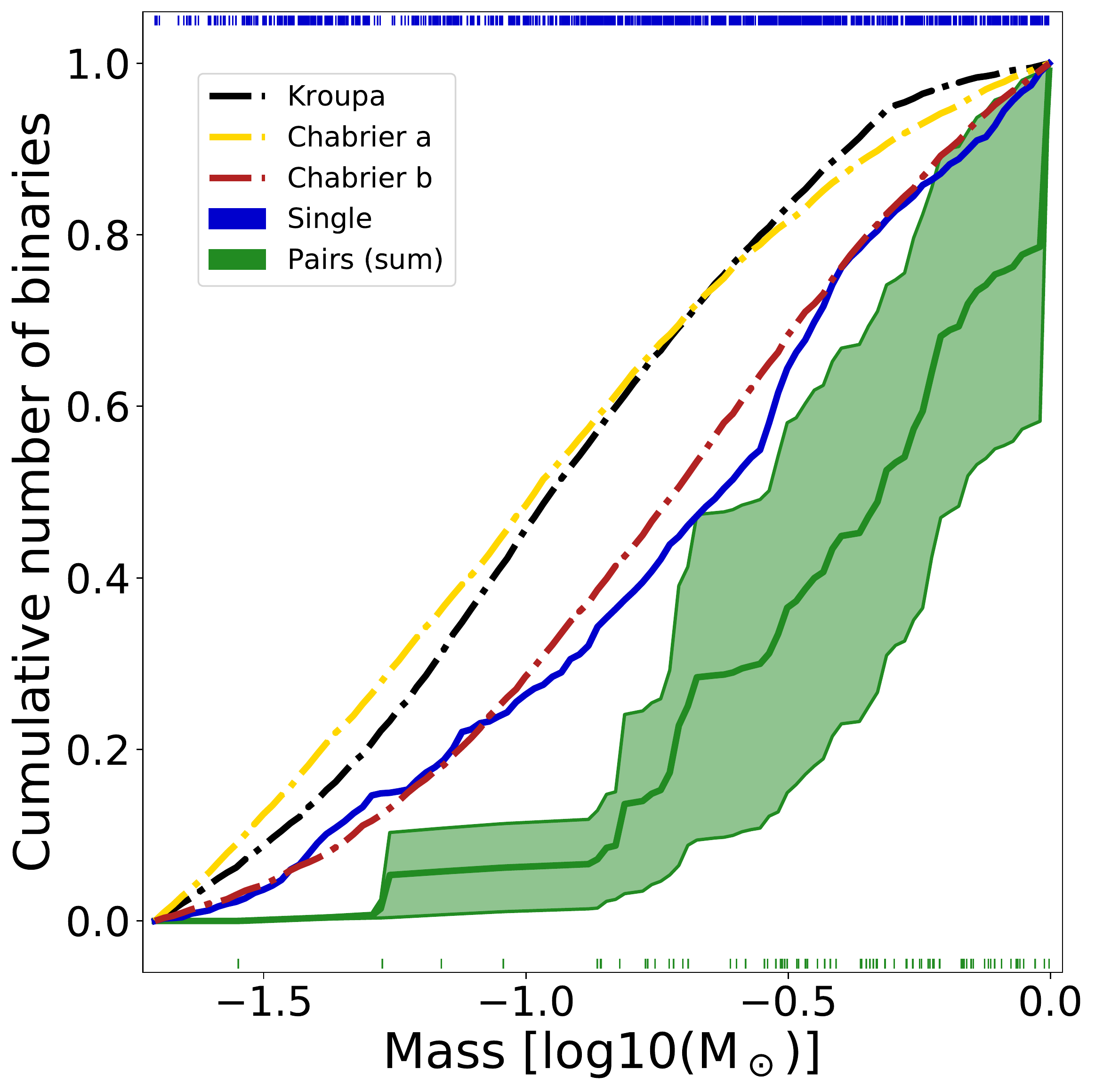}
\caption{Families of curves obtained through the simulations of \emph{completed} cumulative distributions, as explained in the text. Also shown for reference are the cumulative distributions obtain from a Kroupa IMF (dash-dotted black line), a Chabrier IMF for single systems (Chabrier a: dash-dotted yellow line), and a Chabrier IMF with unresolved (Chabrier b: dash-dotted brown). The colored vertical ticks at the top and bottom of the plot show the total mass of each system used to generate the synthetic populations. \label{Fig:CumMass}}
\end{center}
\end{figure*}
In order to probe the Initial Mass Function of multiple systems, in Figure \ref{Fig:HISTmasses} we show the histograms of the primary and companion masses. 
We fit the histograms using broken power laws (i.e. $\sim m^{-\Gamma_i}$), adopting the peak of each specific sample as the breaking point, obtaining the results shown in Table \ref{Tab:Gammas}. 
Even though the values of $\Gamma_1$ are compatible within the errors, both the $\Gamma_2$ and the peak of the two populations is not compatible within $1\sigma$.
To further characterize the possible differences between the mass distributions of primaries and companions and how they compare to the mass distribution of single stars in the ONC, we show in Figure \ref{Fig:CumMass} a set of cumulative mass distributions 
obtained following the same procedure introduced in Section \ref{section:SepOriC}. The top left panel shows the comparison between single systems, primaries and companions. The top right panel shows the comparison between single systems and the full set of masses, both primaries and companions taken individually (we refer to this joint set of mass values as "union"). The bottom panel shows the same comparison where we coadded the mass of the two components of each pair (we refer to this set of mass values as "sum").
In each plot we also show the cumulative distribution obtained from a Kroupa IMF \citep{Kroupa2001}, a Chabrier IMF for single objects \citep[Chabrier a: eq. 17 in][]{Chabrier2003}, a Chabrier IMF with unresolved binaries \citep[Chabrier b:  eq. 18 in][]{Chabrier2003}. 
To avoid introducing biases due to the saturation limit of our survey, we cut the mass distributions at 1~M$_{\Sun}$. As explained in Sec. \ref{section:SepOriC}, we generate one thousand \textit{complete} samples for each population. For each combination we perform a 2-sample KS-test. The results,
summarized in Table \ref{Tab:KS-test}, are characterized by the ratio $n=\frac{n_i}{n_{tot}}$ where $n_i$ is the number of times the KS-test provides a p-value $\leq$ 0.01 (corresponding to a confidence level $>99\%$ that the two population are distinct) and $n_{tot}$ is the total number of simulations. As the ratio increases, it is safer to reject the hypothesis that the two samples are drawn from the same population. The results suggest that the populations are generally different, in particular
a) the mass distribution of the binaries is different from the mass distribution of single stars, b) both are different from the Kroupa/Chabrier IMFs, and c) the primary and companion mass distributions are  different from each other (as already noted in Fig \ref{Fig:HISTmasses}). 
The "union" mass distribution is compatible with a Chabrier IMF with unresolved binaries in $\sim 31\%$ of the tests. The "sum" mass distribution is always incompatible with any Kroupa/Chabrier IMFs.

We interpret these inconsistencies as a result of a systematic deficiency of companion detections below $~100$ AU. Regardless of our best efforts and of our advanced detection techniques, the technical limit of 1-2 pixels for the closest resolvable pairs is basically insurmountable.

Although in this simple exercise we try to enhance the number of binaries by making use of our completeness tests, it must be remarked that the enhancement is only partial. For every detected binary we can compute the chance for that binary to be detected at exactly the separation and magnitude contrast at which it is detected, and we can enhance our sample by one minus that chance.
However,  we cannot account for the truly undetected binaries (i.e., the truly close pairs and those with high flux contrast). A demonstration of this is that our "Single" stars sample (blue line in the top left panel of Figure~\ref{Fig:CumMass} follows the distribution of stellar systems (including unresolved binaries) by \cite{Chabrier2003}, an obvious sign that many binaries are actually hiding within our singles.

For the same reason, even the the conclusions on dissimilarities of the mass distribution of primaries and companions in detected pairs can only be partial, due to biases affecting which systems are preferentially detected as such.

A more complete exercise, involving modeling the a-priori binary mass distribution, SMA, inclinations, eccentricity and spatial distribution within the cluster will be the focus of an upcoming paper in this series (Pueyo et al, in preparation).

\begin{deluxetable*}{cccccccc}
\tablecaption{Comparison between different samples. The table show the ratio $n$ of KS two sample test providing a p-value below 0.01 over the total number of simulations. Bigger n allow us to safely reject the hypothesis that the two samples of each tests are drawn from the same population. \label{Tab:KS-test}}
\tablehead{\colhead{}&\colhead{Kroupa} &\colhead{Chabrier a} & \colhead{Chabrier b}&\colhead{Singles} & \colhead{Companions}}
\decimals
\startdata
 \hline
 \hline
 Primaries     & 1      & 1     & 1     &   1     &  1      \\
 Companions    & 0.94   & 0.92  & 1     &   1     &  -      \\
 Pairs (Union) & 1      & 1     & 0.69  &   0.79  &  -      \\
 Pairs (Sum)   & 1      & 1     & 1     &   1     &  -      \\
\enddata
\end{deluxetable*}


\subsection{Mass ratio}\label{Section:MassRatio}
\begin{figure}[t!]
\begin{center}
\includegraphics[width=0.49\textwidth]{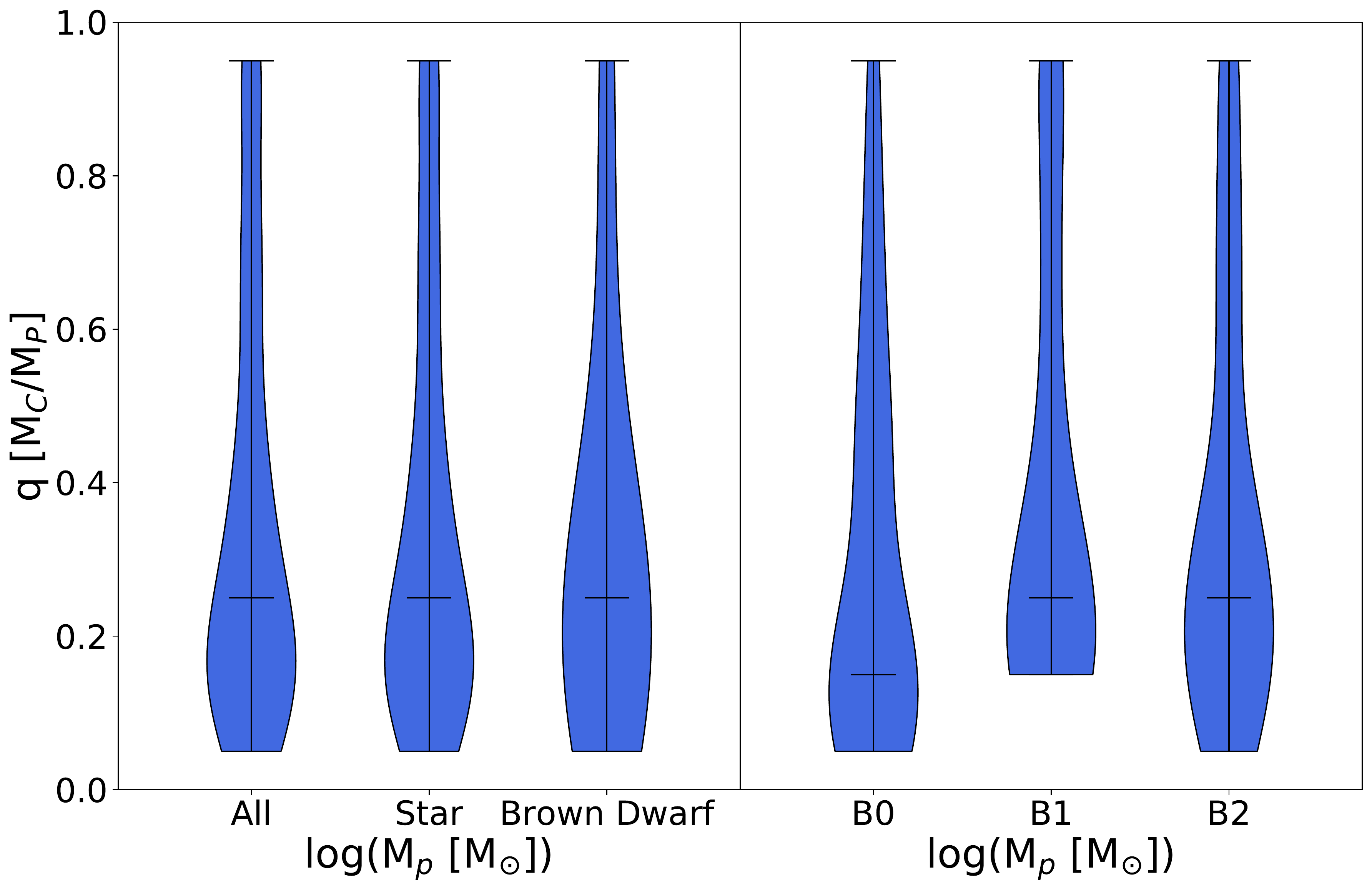}
\caption{Left panel: violin plots of the mass ratio distribution for all candidates and candidates with primaries in the stellar/brown dwarf mass regime. Right panel:same as left panel for different bins of mass of the primary (see the Table \ref{Tab:qdist} for more details). The shape of each distribution show the probability density of the data smoothed by a kernel density estimator while the horizontal black lines mark the median value for each of them  \label{Fig:Vplots}}
\end{center}
\end{figure}
\begin{figure}[t!]
\begin{center}
\includegraphics[width=0.49\textwidth]{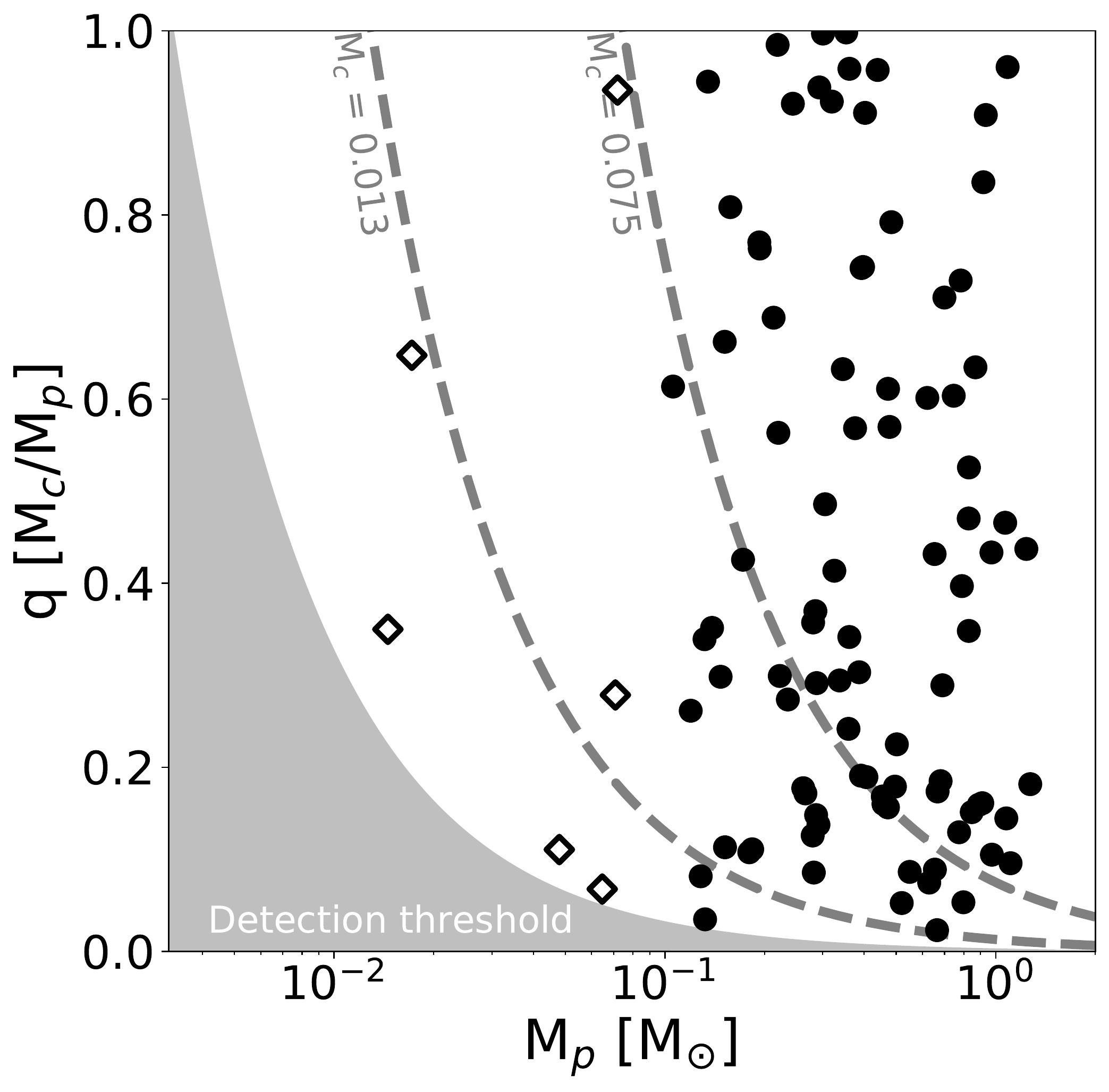}
\caption{Distribution of mas ratio as a function of mass of the primary for the \textit{ONC candidate binary catalog} objects. The primary masses for each candidate are shown by their shape (circle = star; hollow diamond = brown dwarf).The two grey dashed and dot dashed lines show the values of q for which $M_{c} = 0.075$ and $M_{c} = 0.013$ as a function of $M_p$ \label{Fig:Qplots}}
\end{center}
\end{figure}

\begin{deluxetable}{cccrc}
\tablecaption{Median value of q and power-law index $\gamma$ obtain for different range of mass of the primary.\label{Tab:qdist}}
\tablehead{\colhead{Label} & \colhead{q-median} & \colhead{$\gamma$}}
\decimals
\startdata
All  & 0.25 & $-0.7\pm0.2$\\
Star & 0.25 & $-0.7\pm0.2$\\
BD   & 0.15 & $-0.9\pm0.8$\\
\hline
\hline
B0 & 0.15 & $-0.8\pm0.4$\\
B1 & 0.25 & $-0.5\pm0.3$\\
B2 & 0.30 & $-0.6\pm0.3$\\
\enddata
\end{deluxetable}
In this final section we analyze the mass ratio distribution q $=\frac{M_C}{M_P}$, grouping binaries in different bins according to the mass of the primary and following the classification adopted to produce Table~\ref{Tab:MFtable}. The results are shown as violin plots (i.e. a method for graphically depicting groups of numerical data similar to a box plot with a marker for the median of the data and the addition of a rotated kernel density plot on each side).
Overall, we obtain a median value for the mass ratios $q\sim 0.25$, indicating a deficiency of similar-mass binaries (which would have $q\sim 1$). This result is in agreement with what reported by \cite{Duchene2018} for smaller separations (10-60 au).
To compare our results with others work, we characterize the distribution of mass ratio as a power law  $f(q) \propto q^{\gamma}$. Fitting the completeness-corrected histogram, we determine the median values of $q$ and $\gamma$ reported in Table \ref{Tab:qdist}, for the different mass bins.  From a theoretical point of view we would expect that binaries with separation $\lesssim 100$ AU most likely have formed through fragmentation of the protostellar disk while wider systems via free-fall fragmentation during early collapse. Because these two process occur at different times and through different mechanics, it's reasonable to expect them to produce companions with different mass functions and in turn different distribution of mass ratios. We tested this hypothesis obtaining $\gamma_{\lesssim100 AU} = -1.1\pm0.5$ and $\gamma_{\gtrsim100 AU} = -0.6\pm0.2$, finding that the distribution whit separation $\gtrsim 100$ AU (with a bigger and better constrained sample) is incompatible at $2.5\sigma$ from the population of binaries with separation $\lesssim 100$ AU.

\cite{Correia2013} studied eight adaptive optics spatially-resolved binaries in the ONC (along with seven binaries from the literature) in separation range $85-560$ AU and primary mass $0.15-0.8$, finding $\gamma = 1.03 \pm 0.66$, $\gamma = 1.11 \pm 0.37$ and $\gamma = 0.57 \pm 0.38$ for the B98, PS99 and S00 pre-main sequence tracks, respectively. The author find good agreement between their results in the ONC and other star forming regions (e.g. Taurus-Auriga), while our results seems to disagree with both (see below about our comparison with Taurus-Auriga). We think this discrepancy can be explained by the small number of candidates adopted in their survey and in the large amount of close-in small mass companion detected in ours. 
We decided to test this assumption down-sampling the number of candidates in our catalog, randomly extracting the same number as in \citeauthor{Correia2013} in a similar range of masses and separation $\gtrsim 100$ AU. We repeated this operation one hundreds times finding that in $88\%/78\%/38\%$ of the cases we agree within $2\sigma$ with the results from PS99/S00/B98 tracks. It is worth to notice that the candidate we exclude for this test have averege completeness value of $76\%$, and any candidate with completeness smaller than $30\%$ have been detected through multiple visits.
So we conclude that the discrepancy can be attributed to the presence of close in small mass candidate companion we detected through KLIP analysis in our work.

\cite{Kraus2011} conducted a high resolution imaging survey of the Taurus-Auriga star forming region probing the range of separations between $15-5000$ AU, primary and companion masses in the range $0.25-2.5\:M_{\sun}$ and  $0.01-1.17$ M$_{\sun}$, respectively, obtaining $\gamma = 0.2\pm0.2$ at separation $\lesssim 100$ and $0.08 \pm 0.2$ at separations $\gtrsim 100$, i.e. finding an almost flat distribution of $q$ with at most a slight excess of similar mass binaries. Instead, we find an overabundance of low-$q$ binaries. This result still holds even if we consider a range of overlapping primary and companion masses and separations between the two surveys ($0.28-1.27 \:M_{\sun}$ and  $ 0.01-1.04$ M$_{\sun}$ and $ 66-680$ AU respectively), obtaining $\gamma_{Kraus} = 0.3\pm0.3$ and our $\gamma = -0.4\pm0.2$. If instead we limit both dataset at separation $\gtrsim 100 AU$ ad companion masses $\gtrsim 0.05$ M$_{\sun}$, the gamma obtained from the two surveys are now compatible within $\sim1 \sigma$, reconciling the difference. 
\cite{Kraus2011} also remark that their mass-ratio distribution is in stark contrast with \cite{Duquennoy1991}, who studied field binaries with spectral type between F7 to G9 spectral type ($\sim 0.8 - 1.4$  M$_{\Sun}$) and found a mass-ratio distribution peaked towards low masses (q $\sim 0.3$) with few similar mass companions, a finding very close to our result, q$\sim 0.25$. They derived the $\gamma$ from the \cite{Duquennoy1991} dataset, obtaining $\gamma_{q:0-1.1}=-0.36\pm0.07$ and $\gamma_{q:0.2-1.1}=-1.2\pm0.2$. This last value, obtained with a stronger fit -- $\chi_{\nu} =0.7$ with 7 degrees of freedom, is in good agreement with the $\gamma$ we obtain for close in companions (separation $\lesssim 100$ AU) and for primary masses $0.5-1.27$ M$_{\Msun}$ (labeled 'B0' in Table~\ref{Tab:qdist}).
These results, together with the results about the multiplicity fraction presented in Sec.~\ref{Section:BinaryFrequency}, suggest that ONC binaries may represent a template for the typical population of field binaries, upholding the hypothesis that  the ONC may be regarded as a most typical star forming region in the Milky Way.

Figure~\ref{Fig:Qplots} shows the mass ratio of each pair vs. the mass of the primary, i.e. the detailed distribution of the data points used to create Figure~\ref{Fig:Vplots}. The shape of each point indicates the mass of the primary (circle = star; hollow diamond = brown dwarf).
The limits for substellar and planetary mass companions are shown as dashed lines.
The gray area represents the region of parameter space inaccessible  due to our detection limits. Figure~\ref{Fig:Qplots} shows an overabundance of companions around stellar vs. brown dwarf primaries, consistent with the general trend for star forming regions and young associations \citep{Duchene2013}. When detected, very-low mass companions tend to have $q\leq0.4$.
If present, very-low mass binary systems with nearly equal mass must have remained unresolved, with a  projected SMA smaller than our inner separation limit at the distance of the Orion Nebula. In fact, \cite{Winters2019} find the majority of VLM objects in a local volume 25~pc radius have $q\gtrsim0.4$ and their separation peaks at $\sim 20~$AU. As a comparison,  the smallest separation we resolve is $\simeq 50$~AU with low completeness $\mathcal{C} \sim 0.1$. On the other hand, our data seem to suggest that very-low mass binary systems with nearly equal mass and wide separation are exceptionally rare, a possible indication that core fragmentation at the lowest masses favors the formation of asymmetrical systems.


\section{Conclusion}
\label{section:conclusion}
We performed a new analysis of \textit{HST} WFC3/IR images of the Orion Nebula Cluster 
using the Karhunen-Lo\`{e}ve Image Projection (KLIP) algorithm to find faint companions around low-mass primaries. Starting from a sample of 1392 unique bona-fide cluster targets, we find:
\begin{itemize}
    \item 39 candidate binary systems within separation $0.16''-0.77''$ and mass range  M$_p$ $\sim 0.015-1.27$  M$_{\sun}$ for the primary and M$_c$ $\sim 0.004-0.54$  M$_{\sun}$  for the companion.  Of these, 21 are detected for the first time ever. The detection of the $H_2O$ absorption feature allows us to assess with high confidence the membership of these sources in the ONC, although final confirmation of their nature as gravitationally bound systems will require future proper motion studies; 
    \item the overall multiplicity fraction for the ONC determined from the HST/WFC3-IR data, is $11.5\% \pm 0.9\%$. In comparison with other star forming regions, this value is $\sim 2$ times smaller than e.g. Taurus over a similar separation range \citep{Duchene2013}. We find approximately the same binary frequency in the field and in ONC \citep{Duquennoy1991}; 
    \item the mass distribution of the sources belonging to a binary system (either primaries, companions, or combined) is different from the mass distribution of single stars; the primary and companion mass distributions are also different from each other; 
    \item the mass ratio distribution is compatible with what expected from a scenario where close in  binaries formed through fragmentation of the protostellar disk while wider systems formed via free-fall fragmentation; and
    \item an almost complete absence of brown dwarfs and VLM M-dwarfs pairs with similar mass (high-$q$ systems), and a steep distribution of mass ratios peaked towards small $q$-values (median values $q\simeq0.25$).

\end{itemize}
Overall our results suggest that ONC binaries may represent a template for the typical population of field binaries, supporting the hypothesis that the ONC may be regarded as a most typical star forming region in the Milky Way.

\clearpage
\appendix
\section{Receiver operating characteristic curves}
\label{app_section:ROC curves}
A Receiver Operating Characteristic (ROC) curve is a plot that shows the diagnostic ability of a binary classifier system as the discrimination threshold (\textit{T}) varies.  
The ROC curve is created by plotting the true positive rate (or TPR) versus the false positive rate (FPR) at various threshold \textit{T} values. When \textit{T} is set low enough, we accept the whole distribution of TP, but we also accept the whole distribution of FP, so in the ROC curve plot we are at the point (1,1). When we increase \textit{T}, we will lose some TP as well as some FP (the exact rate and so the shape of the ROC curve depend on the exact distribution of the two populations) until we reach the point (0,0) where the selected threshold excludes all the TP and FP. 

To build ROC curves for our detection we first need to simulate the TPR and FPR population representative of each of our candidates. Our sensitivity strongly depends on the magnitude of the primary ($\mathrm{m_{F130M}}$), the contrast ($\Delta\mathrm{mag}$) achieved by PSF subtraction, and the distance of the companion from the primary (separation). We therefore sorted our targets into magnitude bins of the primary from 10 to 22 , $\Delta\mathrm{mag}$ from 0 to 10 (both with a width equal to 1) and separation from $0''$ to $1 ''$ in step of $0.1''$. To build the TPR distribution and the FPR distribution for each of these configurations:
\begin{itemize}
    \item we created one thousand fake binaries. To simulate both the primary and the companion component, we first simulated an isolated star using the model of the PSF obtained from KLIP, re-scaled to match the flux of the object we want to simulate. To perturb the PSF model, we created a local model of the noise combining WFC3 error maps from all the stars of the survey in the same magnitude bin of the simulated star. To take into account different pixel phases we add a small shift ($\leq 0.5$ pixel) to the position of the star. Then we inject the simulated companion in the tile of the simulated primary and add the sky to the final combined tile. During this procedure we also saved the tile of the isolated primary for future analysis.
    \item for each simulation (either the binary or the isolated primary), we perform the same PSF subtraction process illustrated in Section \ref{subsection:PSF subtraction}, retrieving the value of the (positive) signal to noise ratio (SNR) in the pixel where we injected (building the TPR) or did not inject the companion (building the FPR). We decided to use only the positive values to build the ROC curves because by definition the signal from a candidate detection has to be positive.
\end{itemize}

To encapsulate in a single number the performance of our model to distinguish between classifier, we evaluate the Area Under the Curve (AUC) of an ROC. The higher the AUC, better the model is at distinguishing between the true positive population and the false positive population.

\begin{figure*}[h!]
\begin{center}
\includegraphics[width=0.75\textwidth]{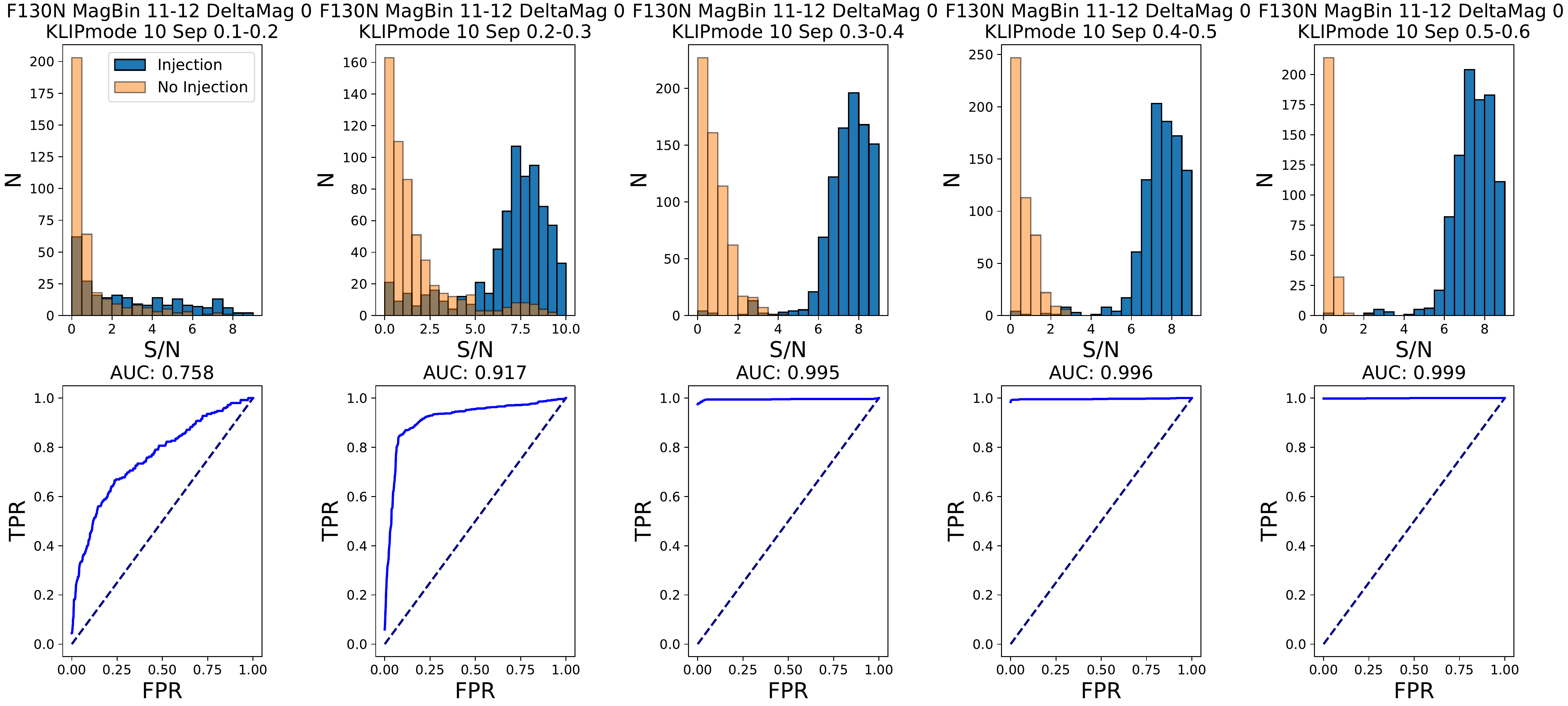}\\
\includegraphics[width=0.75\textwidth]{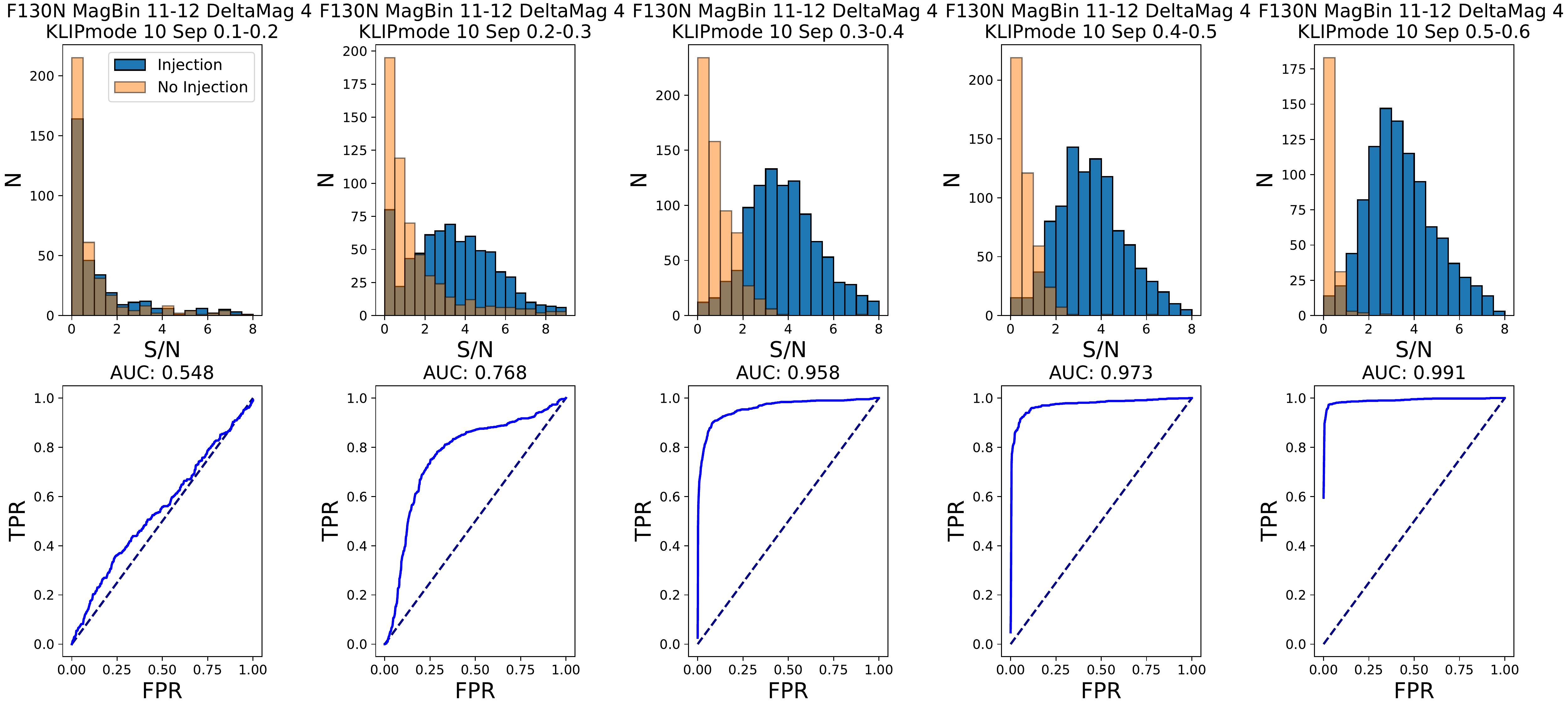}\\
\includegraphics[width=0.75\textwidth]{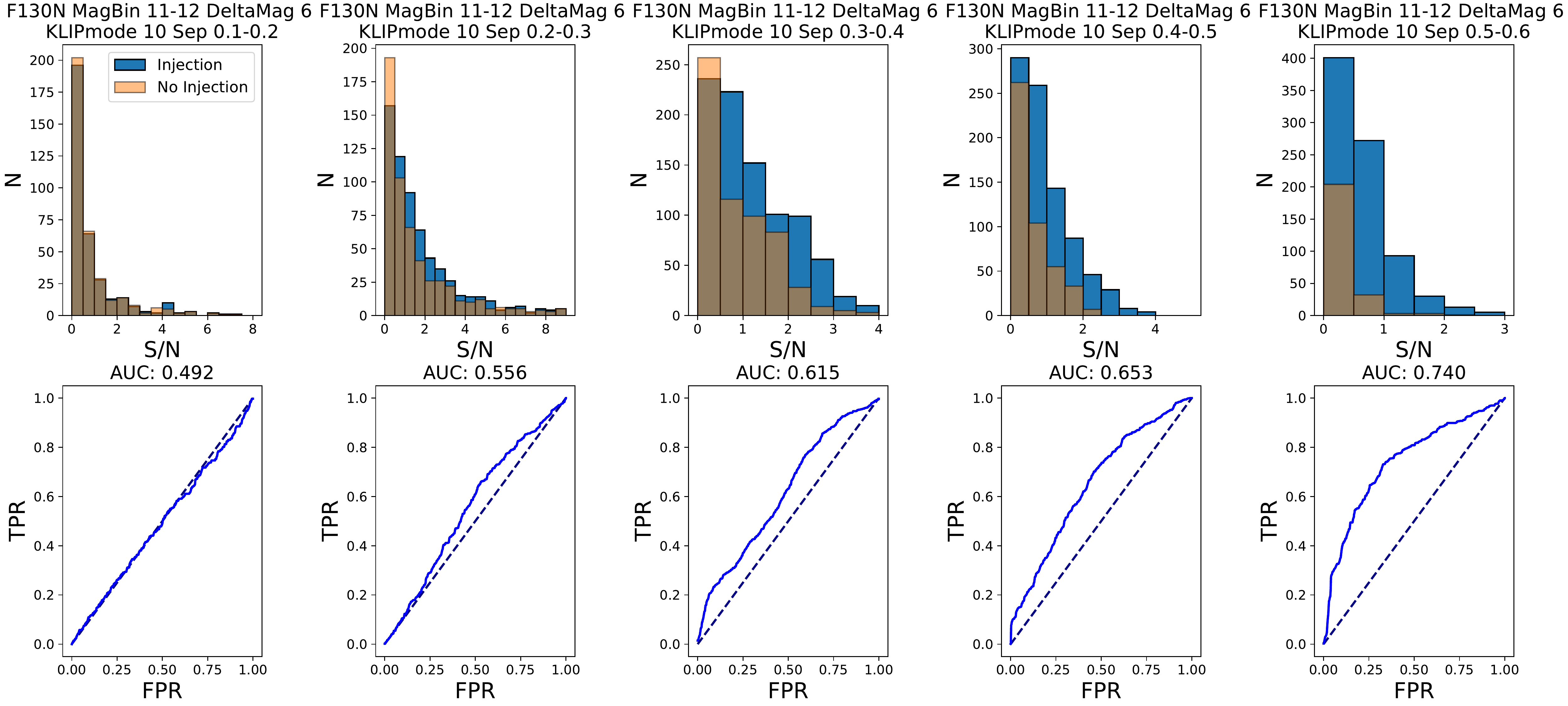}\\
\caption{Distributions of signal to noise and derived ROC curves for filters F130N, magnitude bin of the primary 11-12, $\Delta\mathrm{mag}$ 0,4 and 6 and different distances from the center of the tile. \label{Fig:FINJ_11_0}}
\end{center}
\end{figure*}

Figure \ref{Fig:FINJ_11_0} shows examples of the TP (blue) and FP (orange) histograms for a given binary configuration, and the corresponding ROC curve. Also provided for each ROC curve is the value of the corresponding AUC. 

\FloatBarrier
\section{Gallery of binaries}
\label{app_section:Gallery of binaries}

\begin{figure*}[h!]
\begin{center}
\includegraphics[width=0.95\textwidth]{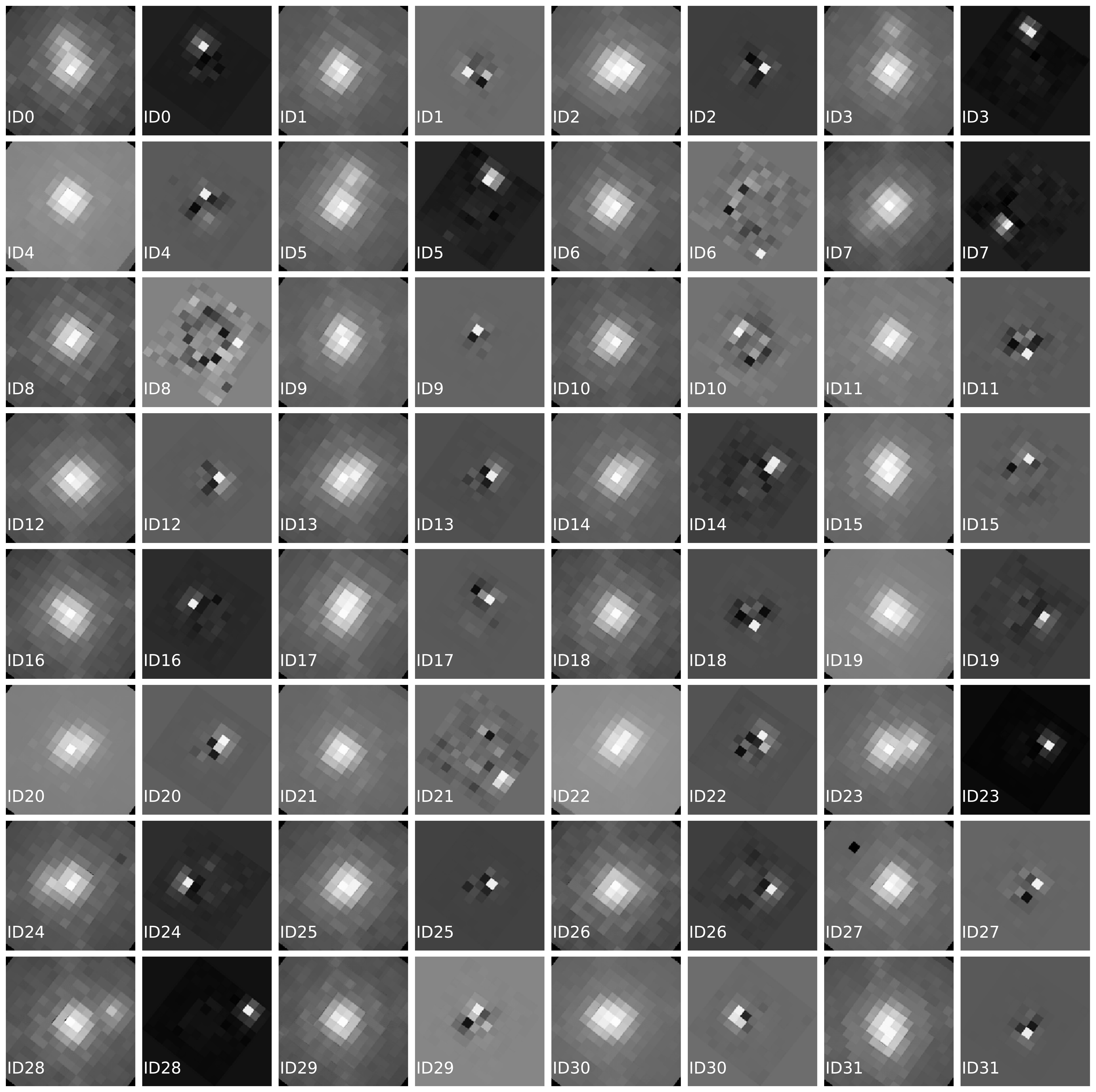}\\
\caption{Each tile shows the residual image after running KLIP for each primary target for which we detect a companion (brighter pixel in the tile). Each stamps has a dimension of 2''x2''. The north is up and east is on the left\label{Fig:KLIP_gallery0}}
\end{center}
\end{figure*}

\begin{figure*}[h!]
\begin{center}
\includegraphics[width=0.95\textwidth]{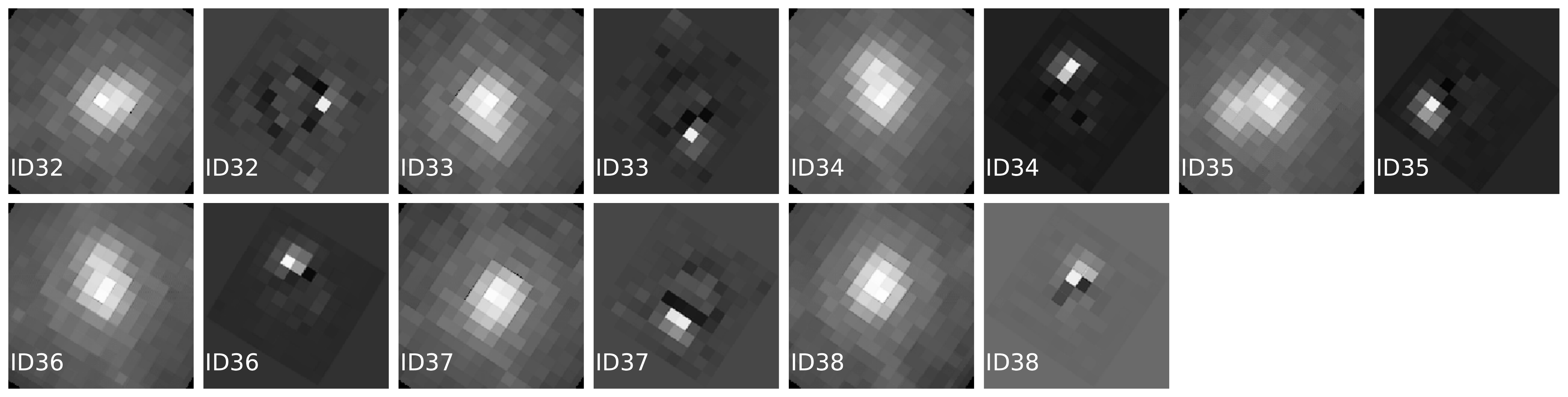}\\
\caption{Each tile shows the residual image after running KLIP for each primary target for which we detect a companion (brighter pixel in the tile). Each stamps has a dimension of 2''x2''. The north is up and east is on the left\label{Fig:KLIP_gallery1}}
\end{center}
\end{figure*}

\begin{figure*}[h!]
\begin{center}
\includegraphics[width=0.95\textwidth]{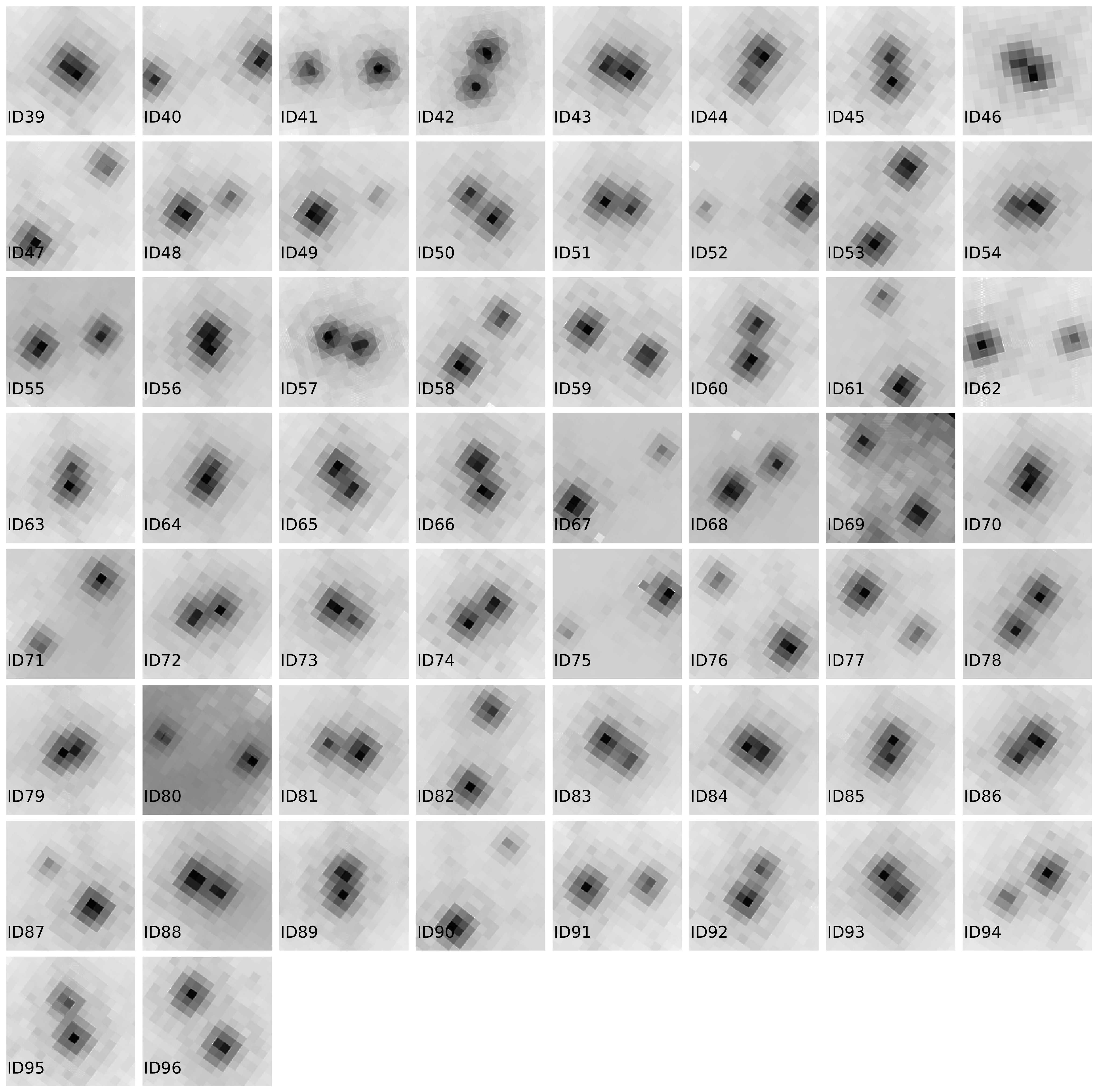}
\caption{WFC3 binaries identified from Catalog~I up to separation $\leq 1.5''$. Each stamps has a dimension of 2''x2''. The north is up and east is on the left \label{Fig:Binary_gallery_close}}
\end{center}
\end{figure*}

Figures~\ref{Fig:KLIP_gallery0} - \ref{Fig:KLIP_gallery1} show the coadded images pre- and post-subtraction for each of the candidate cluster binary presented in Table \ref{Tab:CBC_klip}. Each stamps has a dimension of 2''$\times$2''. 
Figure~\ref{Fig:Binary_gallery_close} shows the postage for the candidates binary from Table \ref{Tab:CBC_close}. Each stamps has a dimension of 2''$\times$2''.
Each postage stamp has been rotated and aligned to have North up and East to the left.

\FloatBarrier
\clearpage

\acknowledgments
The authors thank the anonymous referees for the interesting suggestions and comments.
G. M. Strampelli wants to thanks the Instituto de Astrofísica de Canarias for hospitality.  The authors thank Bo Reipurth for useful comments on the manuscript. Support for Program number GO-13826 was provided by NASA through a grant from the Space Telescope Science Institute, which is operated by the Association of Universities for Research in Astronomy, Incorporated, under NASA contract NASS-26555. CFM acknowledges an ESO fellowship. JA was supported in part by a grant from the National Physical Science Consortium. GMS and AA are supported by the Ministerio de Ciencia, Innovación y Universidades of Spain (grant AYA2017-89841-P) and by the Instituto de Astrofísica de Canarias.
This research has made use of the VizieR catalogue access tool, CDS,  Strasbourg, France. The original description of the VizieR service was published in A\&AS 143, 23. 

\facilities{\textit{HST} (ACS, WFC3)}
\software{Numpy \citep{Numpy}, Astropy \citep{astropy:2013,astropy:2018}, Scipy \citep{SciPy-NMeth:2020}, Matplotlib \citep{Hunter:2007}, PyKLIP \citep{Wang2015}, Pandas \citep{mckinney2010data}}

\clearpage
\bibliography{bibliography} 

\end{document}